\documentclass[pra,superscriptaddress,showkeys,amsmath,amssymb,twocolumn,aps,floatfix]{revtex4-2}
\usepackage[utf8]{inputenc}
\usepackage[T1]{fontenc}
\usepackage{lmodern}
\usepackage{graphicx}  
\usepackage{xcolor}
\usepackage{etoolbox}
\usepackage[colorlinks = true,linkcolor = blue,urlcolor  = blue,citecolor = blue, anchorcolor = blue]{hyperref}
\usepackage{bm}
\usepackage{float}
\usepackage{orcidlink}
\begin{document}

\title{Information-Theoretic Diagnostics of Collective Modes and Vortex Dynamics in Rotating Two-Dimensional Bose-Einstein Condensates}
\author{L.~A.~Machado\orcidlink{0000-0002-7513-7701}}
\affiliation{Instituto de Física de São Carlos, Universidade de São Paulo, CP 369, 13560-970 São Carlos, SP, Brazil.}
\affiliation{School of Mathematics, Statistics and Physics, Newcastle University, Newcastle upon Tyne, NE1 7RU, United Kingdom}
\author{N.~D.~Chavda\orcidlink{0000-0001-5958-1143}}
\affiliation{Department of Applied Physics, Faculty of Technology and Engineering,The Maharaja Sayajirao University of Baroda, Vadodara-390001, India.} 
\author{B.~Chatterjee\orcidlink{0000-0002-1293-4068}}
\affiliation{Department of Physics, Asansol Girls’ College, Asansol, 713304, West Bengal, India}
 \author{M.~A.~Caracanhas\orcidlink{0000-0002-7647-8604} }
 \affiliation{Instituto de Física de São Carlos, Universidade de São Paulo, CP 369, 13560-970 São Carlos, SP, Brazil.}
\author{B.~Chakrabarti\orcidlink{0000-0002-6320-9894}}
\email{barnali.chakrabarti@ufpe.br}
\affiliation{Instituto de Física de São Carlos, Universidade de São Paulo, CP 369, 13560-970 São Carlos, SP, Brazil.}
\affiliation{ Laboratório de Física Teórica e Computacional, Departamento de Física,
Universidade Federal de Pernambuco, 50670-901 Recife, Pernambuco, Brazil}
\author{A.~Gammal\orcidlink{0000-0003-4720-3203}}
\affiliation{Instituto de Física, Universidade de São Paulo, CEP 05508-090, SP, Brazil.}
\author{R.~P.~Sagar\orcidlink{0000-0002-2762-5321}}
\affiliation{Departamento de Química, Universidad Autónoma Metropolitana -
Iztapalapa, San Rafael Atlixco 186, CDMX, 09340, Ciudad de México,
México.}

\begin{abstract}

We investigate the nonequilibrium dynamics of a two-dimensional rotating Bose gas confined in a symmetric anharmonic trap, employing the multiconfigurational time-dependent Hartree method for bosons (MCTDHB). We study states ranging from vortex-free configurations to multicharged (giant) vortices, prepared by tuning the rotation frequency, and analyze their response to sudden interaction and trap quenches. In vortex-free states, interaction quenches induce regular breathing-like dynamics, whereas in the presence of giant vortices they lead to symmetry-breaking surface excitations. In contrast, trap deformations that excite quadrupole-like modes produce stable oscillations in vortex-free condensates but trigger rapid, irregular, and effectively chaotic multimode dynamics in multicharged vortices. To characterize these processes beyond conventional density and phase observables, we employ information-theoretic (IT) measures, including marginal and joint spatial entropies, mutual information, and Kullback--Leibler (KL) divergence, supplemented by an angular-resolved KL measure that quantifies azimuthal density redistribution and symmetry breaking. We find that chaotic splitting
is accompanied by a pronounced growth of information-theoretic indicators, signaling the buildup of
many-body correlations and increasing complexity in the system dynamics. Our results demonstrate the extreme sensitivity of giant vortices to excitation protocols and establish IT measures as quantitative global diagnostics that complement conventional observables by characterizing spatial redistribution and correlations during complex vortex dynamics.

\end{abstract}

\maketitle

\section{Introduction}
The study of nonequilibrium dynamics in ultracold quantum gases provides a unique platform for exploring fundamental aspects of many-body physics, including correlations, coherence, and the emergence of collective excitations~\cite{Bloch2012,Polkovnikov2011,Langen2015,Eisert2015}. Among these systems, rotating Bose--Einstein condensates (BECs) offer a particularly rich setting, as they combine nonequilibrium dynamics with externally controlled angular momentum. Such systems exhibit a wide variety of vortex structures, ranging from the occurrence of quantized vortices~\cite{Matthews1999,Madison2000,Fischer2003a,Chatterjee2024}, giant vortices~\cite{Rougerie2011,Correggi2013,White2024,Tomishiyo_2024}, vortex nucleation~\cite{Raman2001,Dagnino_2009}, quantum Hall effects in rotating condensates~\cite{Cooper2001,Cooper2008}, and the emergence of quantum fluctuations~\cite{Lode:2016,Coddington_2003,Schweikhard_2004}. The stability and dynamics of these vortices are highly sensitive to rotation frequency and external perturbations, making them an ideal system to probe the interplay between geometry, interactions, and excitations~\cite{Kasamatsu2003,Aftalion2005,Sinha2001,Recati2001,Roy_2025,Machado_2025}.

Building on the vortex dynamics discussed above, the role of external confinement is crucial in determining the behavior of rotating quantum gases. Different trapping geometries provide powerful means to control vortex formation, stability, and superfluid flow. In particular, anharmonic trapping potentials have enabled the observation and stabilization of superfluid currents in ultracold atomic systems~\cite{Bretin2004,Stock2005}. Ring-shaped (toroidal) traps, on the other hand, offer an ideal platform for generating and sustaining persistent currents in Bose--Einstein condensates~\cite{Ryu2007}. From a theoretical perspective, a wide range of studies have explored rotating condensates in more complex settings, including fully three-dimensional configurations~\cite{AboShaeer2001,Fetter2009} as well as quasi-two-dimensional systems confined in asymmetric, pancake-shaped quartic--quadratic potentials~\cite{Huang2010,  Fetter2001,Kasamatsu2002,Brito}.

A substantial body of work has investigated rotating Bose–Einstein condensates within the framework of the Gross–Pitaevskii (mean-field) approximation, revealing a wide range of phenomena such as vortex formation, lattice structures, and collective excitations~\cite{Fetter2009,Sinha2001,Fetter2001,Kasamatsu2003}. However, this approach inherently neglects many-body correlations and quantum fluctuations beyond the mean-field description. In contrast, comparatively fewer studies have explored rotating condensates within a full many-body framework~\cite{Cooper2001,Wilkin1998,Bertsch1999,Jackson2001,Viefers_2000,Cremon_2013,Cremon_2015,Molignini_2025,molignini2026stabilitydecaymacrovorticesrotating,Machado_2025,Roy_2025,Dutta_2023,Weiner_2017}, where intriguing effects such as condensate fragmentation and the build-up of correlations can emerge. These features are crucial for understanding regimes where the mean-field approximation breaks down, particularly in strongly interacting or highly excited systems.

Motivated by these considerations, we investigate a two-dimensional rotating Bose gas confined in a symmetric anharmonic trap and analyze its response to sudden quenches that excite collective modes. The choice of a symmetric anharmonic trap allows access to high-rotation regimes where purely harmonic confinement becomes unstable, thereby enabling the formation and stabilization of multicharged (giant) vortices while preserving rotational symmetry. By tuning the rotation frequency, the system exhibits a crossover from a vortex-free regime to states hosting quantized vortices with few charges, as well as multicharged (giant) vortices. We consider two excitation protocols: (i) interaction quenches and (ii) trap deformations, providing complementary routes to probe collective dynamics in the system. To capture the underlying dynamics, we employ the multiconfigurational time-dependent Hartree method for bosons (MCTDHB)~\cite{Alon:2007,Alon:2008,Streltsov:2006,Streltsov:2007,Lode:2016} implemented in MCTDHX-software~\cite{Lin:2020,MCTDHX}, which provides a fully correlated many-body description and enables us to investigate the nonequilibrium response of the system following sudden quenches.

To characterize the quenched dynamics beyond standard density and phase
observables, we employ information-theoretic (IT) measures~\cite{CoverThomas2006,Laguna_2011,Peng_2015,Angulo_2022,Schurger2024},
including the marginal and joint entropies $(S_x,S_y,S_{xy})$ and the
mutual information $I$. While density and phase profiles directly
characterize vortex structure, angular Fourier components resolve
specific symmetry-breaking modes, and natural-orbital occupations
quantify many-body fragmentation, the IT measures provide complementary
global measures of spatial redistribution and correlations. In
particular, $S_{xy}$ quantifies the overall spatial spreading of the
density, whereas the mutual information $I=S_x+S_y-S_{xy}$ isolates
correlations between the two spatial directions. We further employ an
angular-resolved Kullback--Leibler (KL) divergence to quantify the
redistribution of angular probability associated with symmetry breaking.
Its temporal behavior can be directly related to angular Fourier
components such as $A_2$ and $A_4$, while providing an information-based
measure of the corresponding angular redistribution. Thus, the IT
measures are not intended to replace conventional vortex observables, but
to provide complementary quantitative diagnostics of global spatial
redistribution and correlations.

Our results reveal a strong dependence of vortex dynamics on both the
excitation protocol and vortex charge. Weaker interaction quenches induce
regular breathing-like dynamics in vortex-free states and only weakly
perturbed dynamics in multicharged (giant) vortices, with little spatial
density deformation and weak many-body fragmentation. In contrast,
stronger quenches drive symmetry-breaking vortex splitting accompanied
by pronounced spatial density deformation, enhanced many-body
fragmentation, and the buildup of many-body correlations.

Trap deformations, on the other hand, produce regular oscillations in
vortex-free condensates but trigger rapid and irregular splitting
dynamics in multicharged vortices. This behavior is driven by the
intrinsic instability of giant vortices, giving rise to complex,
seemingly chaotic dynamics even in the absence of pronounced
many-body fragmentation. Consistently, IT
measures exhibit strongly irregular temporal fluctuations without a
sustained overall growth, indicating that the dynamical complexity is
not necessarily accompanied by a substantial redistribution of
natural-orbital occupations. Thus, the IT measures provide quantitative
diagnostics of the complex spatial dynamics even when many-body
fragmentation remains relatively weak.

The present work addresses three central questions: how different
excitation protocols influence the stability of giant vortices, how the
onset of complex and chaotic dynamics manifests in IT measures, and to
what extent these measures provide a quantitative framework for
characterizing nonequilibrium behavior in rotating quantum gases.

Our results demonstrate that the stability of multicharged vortices is
governed by a subtle interplay between rotation, trap geometry, and
many-body correlations. By combining high-resolution many-body
simulations with conventional vortex observables and IT diagnostics, we
identify distinct dynamical responses to interaction and trap quenches,
including regular collective oscillations and irregular vortex-splitting
dynamics. The IT measures provide complementary global diagnostics of
spatial redistribution and correlation buildup, remaining sensitive to
the dynamical complexity even when local density observables become
strongly depleted or many-body fragmentation remains relatively weak.
This combined approach provides a quantitative framework for
characterizing complex nonequilibrium vortex dynamics and offers a
pathway to investigate the limits of mean-field descriptions in
strongly rotating Bose--Einstein condensates.

It is important to note that, the information-theoretic measures are employed
to characterize the spatial redistribution of the one-body density
and the statistical dependence between the spatial coordinates. We emphasize that
they do not, by themselves, quantify many-body fragmentation,
which is instead identified through the occupations of the
natural orbitals.

The remainder of this paper is organized as follows. In Sec.~\ref{sec2}, we present the model Hamiltonian and the quench protocol, methodology and observables in different subsections. In Sec.~\ref{sec3}, we present the results, which are subdivided into several sections. Sect.~\ref{sec4} presents the conclusions.

\section{Model, Methodology, and Observables}
\label{sec2}

\subsection{Model Hamiltonian and quench protocol}
\label{sec:model_hamiltonian}

We consider a two-dimensional rotating Bose–Einstein condensate (BEC) of $N$ identical bosons confined in a symmetric anharmonic trap. The nonequilibrium dynamics following an interaction quench is obtained by solving the time-dependent Schr\"odinger equation,
\begin{equation}
\hat{H} \Psi(\mathbf{r}_1,\dots,\mathbf{r}_N,t) = i\hbar \frac{\partial}{\partial t} \Psi(\mathbf{r}_1,\dots,\mathbf{r}_N,t).
\end{equation}

In the rotating frame, the many-body Hamiltonian is given by
\begin{equation} 
\hat{H} = \sum_{i=1}^{N} \hat{h}(\mathbf{r}_i) + \sum_{i<j} \hat{W}(|\mathbf{r}_i - \mathbf{r}_j|),
\label{propagation_eq}
\end{equation}
where $\hat{h}(\mathbf{r})$ denotes the single-particle Hamiltonian and $\hat{W}$ is the two-body interaction potential.

The one-body Hamiltonian in the rotating frame reads
\begin{eqnarray}
\hat{h}(\mathbf{r}) &=& -\frac{\hbar^2}{2m} \nabla^2 - \Omega \hat{L}_z + V_{\mathrm{pot}}(\mathbf{r}),
\end{eqnarray}
where $\nabla^2 = \partial_x^2 + \partial_y^2$ is the Laplacian, $\Omega$ is the rotation frequency, and $\hat{L}_z$ is the $z$-component of the angular momentum operator,
\begin{equation}
\hat{L}_z = -i\hbar \left( x \frac{\partial}{\partial y} - y \frac{\partial}{\partial x} \right).
\end{equation}

The external confinement is provided by a symmetric anharmonic (quartic) trapping potential of the form
\begin{equation}
V_{\mathrm{pot}}(\mathbf{r}) = \frac{\kappa}{4} \left( x^2 + y^2 \right)^2,
\end{equation}
where $\kappa$ determines the strength of the anharmonic confinement. We choose $\kappa=1$.

The interparticle interaction is modeled by a finite-range Gaussian potential,
\begin{equation}
\hat{W}(|\mathbf{r}_i - \mathbf{r}_j|) 
= g \, \frac{1}{2\pi\sigma^2} 
\exp\left( -\frac{|\mathbf{r}_i - \mathbf{r}_j|^2}{2\sigma^2} \right),
\end{equation}
where $g$ denotes the interaction strength and the mean-field interaction parameter is defined as $\Lambda=g(N-1)$. $\sigma$ characterizes the range of the interaction, we keep $\sigma=0.25$~\cite{Dutta_2023}.

In the limit $\sigma \to 0$, the interaction reduces to a contact (delta) potential,
\begin{equation}
\hat{W}(|\mathbf{r}_i - \mathbf{r}_j|) 
= g \, \delta(\mathbf{r}_i - \mathbf{r}_j),
\end{equation}
which is commonly used in mean-field descriptions of dilute Bose gases.
The finite-range form ensures numerical stability and provides a controlled regularization of the contact interaction in two dimensions.

We solve the dimensionless Schr\"odinger equation obtained by scaling
the Hamiltonian with the natural length scale associated with the
quartic confinement. For
$V_{\mathrm{pot}}(r)=\kappa r^4/4$, we define
$l_0=(\hbar^2/m\kappa)^{1/6}$ and
$\epsilon_0=\hbar^2/(m l_0^2)$. We then adopt natural units
$\hbar=m=\kappa=1$, for which $l_0=\epsilon_0=t_0=1$, with
$t_0=\hbar/\epsilon_0$. All quantities reported below are expressed
in these dimensionless units.

We prepare the initial many-body state of $N=8$ bosons with interaction strength $\Lambda = 0.1$ as the ground state of the Hamiltonian at a given rotation frequency $\Omega$. Depending on the chosen parameters, the system realizes (i) a vortex-free rotating condensate, (ii) a state with a low-charged vortex, or (iii) a multiply quantized (multicharged) vortex state localized at the trap center.

To investigate the nonequilibrium dynamics, we employ two distinct quench
protocols: \\

\emph{(i) Interaction quench:} The system is driven out of equilibrium
by a sudden change in the interaction strength, $g\rightarrow g'$,
at $t=0$. Since the post-quench Hamiltonian remains rotationally
symmetric, a weak interaction quench predominantly induces isotropic
radial dynamics of the condensate, whereas a stronger quench can drive
symmetry-breaking spatial density splitting.\\

\emph{(ii) Trap quench:} The dynamics are initiated by a sudden
deformation of the trapping potential. Specifically, at $t=0$, the
initially symmetric potential is instantaneously made anisotropic by
introducing an elongation along the $x$ direction,
$V_{\mathrm{pot}}(x,y)\rightarrow V'_{\mathrm{pot}}(x,y)$, with
\begin{equation}
V'_{\mathrm{pot}}(x,y)=\frac{1}{4}(0.8x^2+y^2)^2.
\end{equation}
This quench explicitly breaks rotational symmetry and induces anisotropic
shape oscillations of the condensate.

\subsection{Methodology}
Within the multiconfigurational time-dependent Hartree method for bosons (MCTDHB), the many-body wave function of $N$ interacting bosons is expanded in a time-dependent basis of permanents,
\begin{equation}
|\Psi(t)\rangle = \sum_{\vec{n}} C_{\vec{n}}(t)\, |\vec{n}; t\rangle,
\end{equation}
where $\vec{n} = (n_1, n_2, \dots, n_M)$ denotes a configuration specifying the occupations of $M$ single-particle orbitals, with $\sum_{j=1}^{M} n_j = N$. Each permanent $|\vec{n}; t\rangle$ represents a symmetrized many-body state of bosons distributed over the time-dependent orbitals~\cite{Lode:2016,Lode:2020,Alon:2007,Streltsov:2007}.

A key feature of MCTDHB is that both the expansion coefficients $C_{\vec{n}}(t)$ and the single-particle orbitals are time dependent, allowing the underlying basis to adapt dynamically to the evolving many-body state. This time-dependent variational optimization leads to an efficient representation of the system, capturing correlations with a significantly reduced basis compared to fixed-orbital approaches. In the formal limit $M \rightarrow \infty$, the method becomes exact, while in practice a finite number of orbitals is employed to achieve a balance between computational cost and accuracy.

The equations of motion for the coefficients and orbitals are obtained by applying the time-dependent variational principle~\cite{variational1,variational3,variational4}, resulting in a set of coupled nonlinear integrodifferential equations. These equations are solved numerically using the MCTDH-X package~\cite{MCTDHX,Lin:2020}, which provides an accurate and efficient framework for simulating the time evolution of interacting bosonic systems.

\subsection{Observables}

To characterize the nonequilibrium dynamics and vortex evolution, we compute a set of observables probing the density distribution, collective excitations, angular symmetry, and IT  properties. \\

{\emph {One-body density and phase}}\\

We analyze the reduced one-body density matrix $\rho^{(1)}(\mathbf{r},\mathbf{r}';t)$ and its diagonal, the particle density,
\begin{equation}
n(x,y,t) = \rho^{(1)}(x,y; x,y; t).
\end{equation}

The phase profile $\theta(x,y)$ is extracted from the argument of the dominant natural orbital $\phi_1(x,y)$ obtained from the one-body reduced density matrix, i.e., $\theta(x,y) = \arg[\phi_1(x,y)]$. Vortices are identified as phase singularities located at density minima, around which the phase exhibits quantized winding. Specifically, the circulation of the phase along a closed contour enclosing the vortex core satisfies $\left[
\oint \nabla \theta \cdot d\mathbf{l} = 2\pi \ell,
\right] $
where $\ell$ denotes the vortex charge.
\\

{\emph{Natural orbitals}}\\

Diagonalization of \(\rho^{(1)}(\mathbf{r},\mathbf{r}')\) yields the natural orbitals \(\{\phi_j(\mathbf{r})\}\) and their corresponding natural occupations \(\{n_j\}\):
\begin{equation}
\rho^{(1)}(\mathbf{r},\mathbf{r}') = \sum_j n_j \, \phi_j(\mathbf{r}) \phi_j^*(\mathbf{r}').
\end{equation}
Here, \(\phi_j\) denotes the \(j\)-th natural orbital, while \(n_j\) denotes its corresponding natural occupation. We use this notation consistently throughout the manuscript.

\vskip 4pt

{\emph{Collective modes}}\\

The monopole and quadrupole modes are defined as
\begin{equation}
O_{\mathrm{M/Q}}(t)=
\int n(x,y,t)\left(x^2 \pm y^2\right)\,dx\,dy,
\end{equation}
where the $+$ and $-$ signs correspond to the monopole and quadrupole
modes, respectively. The monopole mode characterizes isotropic radial
expansion and contraction, whereas the quadrupole mode measures
anisotropic deformation along the principal axes.

\vskip 4pt
{\emph{Angular Fourier and fourfold operators}}\\

Angular Fourier components are defined by
\begin{equation}
 A_m(t)= \int \; \int n(x,y,t) \left( \frac{x+iy}{r}\right)^{m} \; dx \, dy
\end{equation}
which quantify $m$-fold angular symmetry, with $m=2$ and $m=4$
characterizing two- and fourfold structures, respectively. In
particular, $A_4$ probes the fourfold deformation associated with
giant-vortex splitting. The radially weighted fourfold operator
\begin{equation}
    O_4(t) = \int n(x,y,t) \left( x^{4} - 6 x^{2}y^{2} +y^{4}\right) \;dx \, dy
\end{equation}
enhances sensitivity to symmetry breaking in the outer regions where
vortex fragments emerge.

\vskip 4 pt

{\emph{Shannon entropies and mutual information}}\\

For normalized density $n(x,y,t)$, the spatial Shannon entropy and its
marginal entropies are
\begin{equation}
\begin{aligned}
S_{xy}(t) & = - \int dx \, dy \; n(x,y,t)\,\ln n(x,y,t), \\
S_{\mu}(t) & =-\int n_{\mu}(\mu,t)\ln n_{\mu}(\mu,t)\,d\mu,
\quad \mu=x,y,
\end{aligned}
\end{equation}
where
\begin{equation}
n_x(x,t)=\int n(x,y,t)\,dy,\qquad
n_y(y,t)=\int n(x,y,t)\,dx.
\end{equation}
All densities are normalized to unity. \\

The mutual information,
\begin{equation}
\begin{aligned}
I(t) &= \int dx \, dy \; n(x,y,t)\,
\ln\!\left[\frac{n(x,y,t)}{n_x(x,t)\,n_y(y,t)}\right] \\
     &= S_x(t) + S_y(t) - S_{xy}(t) \geq 0,
\end{aligned}
\end{equation}
quantifies the correlations between the $x$ and $y$ degrees of freedom.
Larger $S_{xy}$ indicates greater spatial delocalization, while larger
$I$ indicates stronger interdependence between the two spatial degrees
of freedom.

\vskip 4pt
{\emph{Kullback--Leibler divergence and angular KL measure}}\\

The Kullback--Leibler (KL) divergence quantifies the deviation of the
evolving density from a reference distribution $n_{\mathrm{ref}}$,
\begin{equation}
D_{\mathrm{KL}}(t)=
\int dx\,dy\,n(x,y,t)
\ln\!\left[\frac{n(x,y,t)}
{n_{\mathrm{ref}}(x,y)}\right]\geq0.
\end{equation}

To resolve angular symmetry breaking, we define the normalized angular
density over a narrow ring centered at $r_0$,
\begin{equation}
P(\theta,t)=
\int_{r_0-\Delta r}^{r_0+\Delta r}
n(r,\theta,t)\,r\,dr,
\qquad
\int_0^{2\pi}P(\theta,t)\,d\theta=1,
\end{equation}
and the corresponding angular KL divergence,
\begin{equation}
D_{\mathrm{KL}}^\theta(t)=
\int_0^{2\pi}P(\theta,t)
\ln\!\left[\frac{P(\theta,t)}{P(\theta,0)}\right]d\theta.
\end{equation}
Here, $D_{\mathrm{KL}}$ measures the global deviation from the reference
density, while $D_{\mathrm{KL}}^\theta$ quantifies changes in the angular
density distribution and hence azimuthal symmetry breaking.

It is important to distinguish the information-theoretic
characterization of the spatial density from the characterization
of many-body fragmentation. The entropies and mutual information
considered here are constructed from the one-body density
distribution and quantify its spatial spreading and the statistical
dependence between the spatial coordinates. They do not directly
measure the occupation of different many-body natural orbitals.
Throughout this work, we therefore use the term spatial
redistribution or density deformation when referring to changes
in the one-body density, reserving the term many-body fragmentation
for situations in which more than one natural orbital acquires
a significant occupation.

%

\section{Results}
\label{sec3}
We now present the nonequilibrium dynamics of the two-dimensional rotating Bose–Einstein condensate following the quench protocols. Our analysis is organized in three parts. First, we characterize the initial states prepared in the system, including the vortex-free state, lower-charge vortex, and multicharged vortex configurations, focusing on their density and phase structure. Second, we study the condensate’s response to an interaction quench, which predominantly excites the isotropic monopole (breathing) mode, including some angular deformation for stronger interaction quench. Finally, we analyze the dynamics following a trap quench that excites the quadrupolar  mode, inducing anisotropic deformations and angular symmetry breaking for giant vortices. Throughout, we employ the observables defined in Sec.~\ref{sec2}—collective mode operators, angular Fourier components, radially weighted operators, and IT measures—to quantify density redistribution, symmetry breaking, and vortex splitting.

\subsection{Initial States}
\begin{figure}[htbp]
    \centering
    \includegraphics[width=\columnwidth]{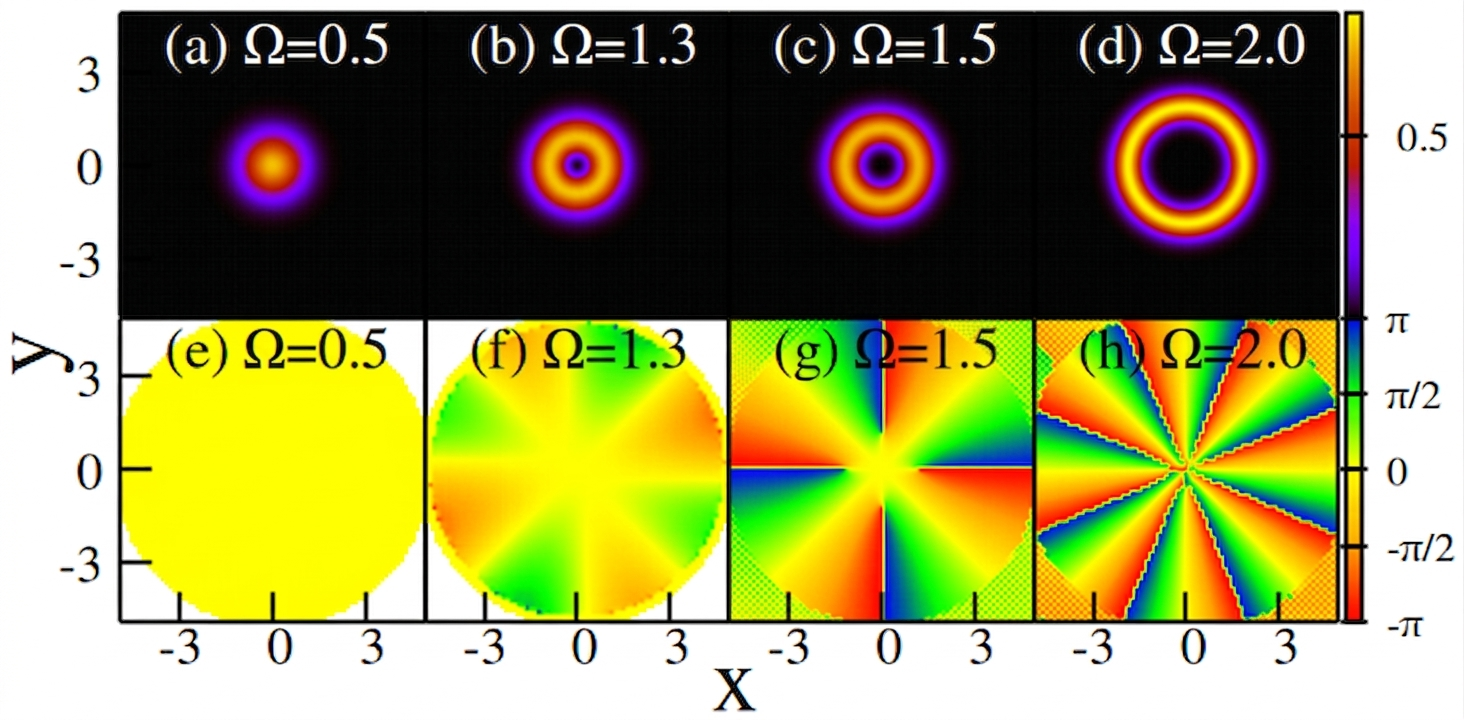}
    \caption{Initial condensate states at selected rotation frequencies $\Omega$. Top panels show the density $n(x,y;0)$, and bottom panels display the corresponding phase. For $\Omega \le 1.2$, the condensate forms a uniform central cloud with no vortex, and the phase is essentially uniform. At $\Omega=1.3$, a central vortex emerges, while $\Omega=1.5$ and $\Omega=2.0$ show increasing vortex charge, evolving into a giant vortex with annular density and a phase indicating total charges of 4 and 8, respectively. Computation is done with $M=6$.}
    \label{fig:initial-selective}
\end{figure}

The initial state of the condensate is shown in Fig.~\ref{fig:initial-selective}, where the top panels display the density and the bottom panels show the corresponding phase. The figure presents a few representative rotation frequencies to illustrate the key features of vortex formation and evolution, while a full set of cases is not shown.

For low rotation, $\Omega \le 1.2$, the condensate forms a nearly uniform central cloud with no vortex. The corresponding phase is essentially uniform, indicating the absence of any phase singularity. At $\Omega=1.3$, a central vortex emerges, manifested as a depletion in the density at the center and a corresponding $2\pi$ phase winding around the core. Increasing the rotation to $\Omega=1.5$, the central vortex develops a clear charge of 4, as seen from the phase structure, while for $\Omega=2.0$ the phase indicates an effective total charge of 8, corresponding to a giant vortex.

The density transitions from a simply connected central cloud at low rotation to an annular structure at higher $\Omega$, reflecting the combined effects of centrifugal forces and the increasing vortex charge. Despite showing only selected cases, the figure captures the qualitative evolution of the condensate topology and the growth of vortex charge with rotation.

\begin{figure}[htbp]
    \centering
    \includegraphics[width=\columnwidth]{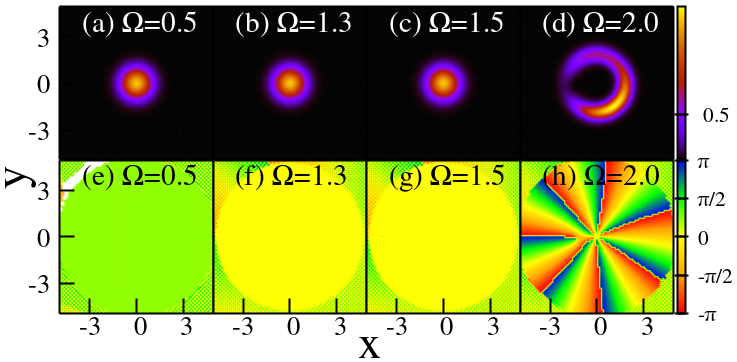}
\caption{Initial condensate states at selected rotation frequencies $\Omega$ obtained within the mean-field approach ($M=1$). Top panels show the density $n(x,y;0)$, and bottom panels display the corresponding phase. For $\Omega \leq 1.5$, the system remains in a condensed state, with the vortex structure evolving as the rotation frequency increases. At $\Omega=2.0$, in contrast to the $M=6$, MCTDHB result, the density profile becomes distorted and does not develop a well-defined annular giant-vortex structure.}
    \label{fig:initial-selective-meanfield}
\end{figure}

To elucidate the role of many-body correlations beyond the mean-field description, we next compare the MCTDHB results with the corresponding Gross–Pitaevskii ($M=1$) calculations. The density and phase profiles obtained for the same set of rotation frequencies are presented in Fig.~\ref{fig:initial-selective-meanfield}. While both approaches yield a condensed state at low rotation frequencies, their predictions increasingly diverge as $\Omega$ is increased. The $M=6$, MCTDHB calculation exhibits the progressive formation of a vortex core, which evolves into a well-developed giant-vortex state at $\Omega=2.0$, characterized by a pronounced annular density profile and the associated phase winding. In contrast, the $M=1$, mean-field description does not produce a stable giant-vortex configuration at the same rotation frequency, instead yielding a distorted density profile without a well-developed annular structure. This qualitative difference indicates that many-body correlations beyond the single-orbital mean-field description play an important role in the emergence and stabilization of the giant-vortex state.

\begin{figure}[htbp]
    \centering
    \includegraphics[width=\columnwidth]{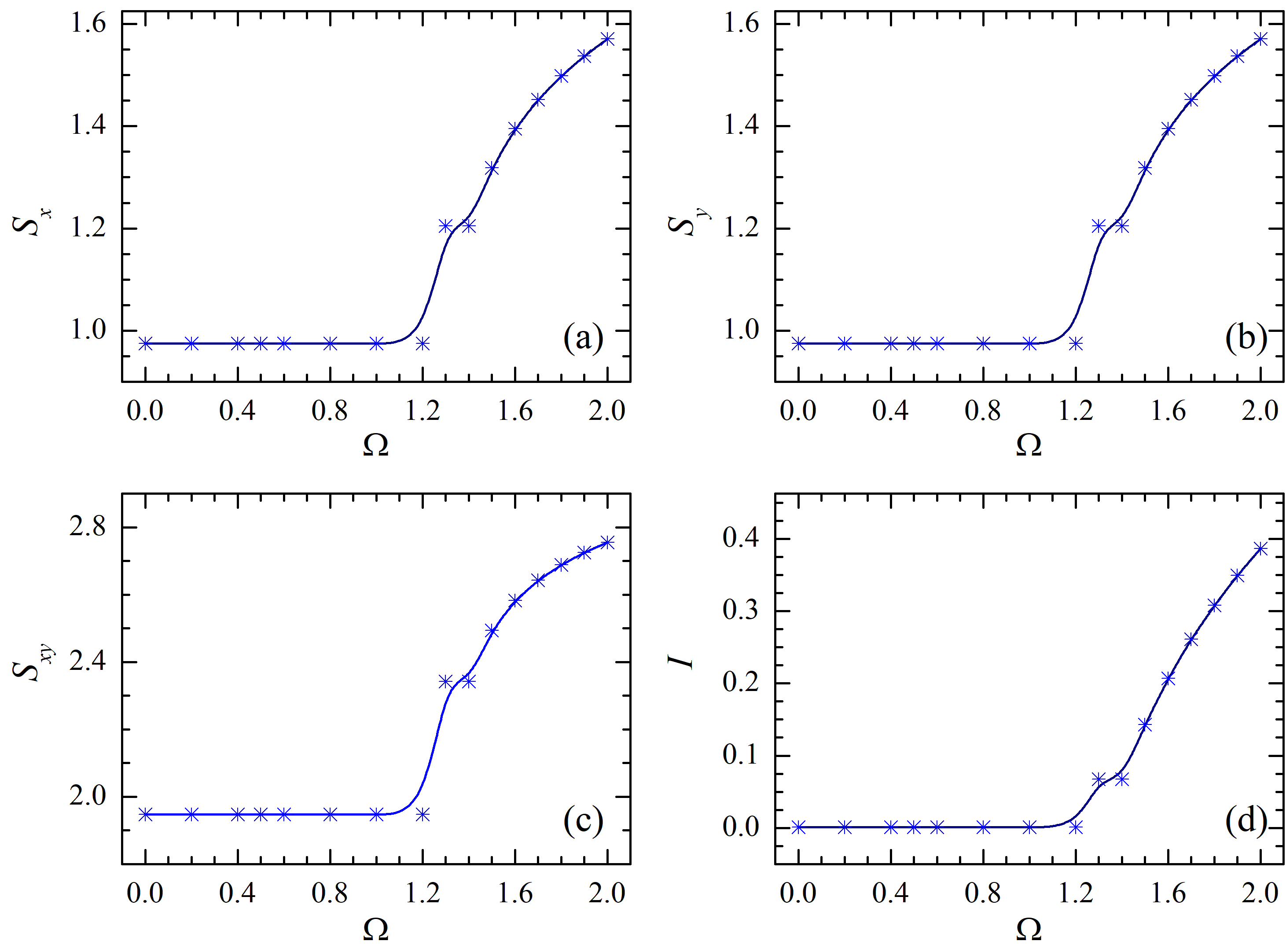}
    \caption{Shannon information entropies and mutual information as a function of rotation frequency $\Omega$. All four quantities, $S_x$, $S_y$, $S_{xy}$, and mutual information $I$, remain constant for $\Omega \le 1.2$. Beyond $\Omega=1.2$, each measure exhibits a kink and increases in a nonlinear manner, signaling the onset of central vortex formation and the emergence of correlations not directly visible in the density or phase profiles.}
    \label{fig:initial-info}
\end{figure}

To quantify the spatial redistribution and correlations beyond the direct
information provided by the density and phase, we calculate the Shannon
entropies $S_x$, $S_y$, and $S_{xy}$, together with the mutual information
$I$, as functions of the rotation frequency $\Omega$ (Fig.~\ref{fig:initial-info}). For
$\Omega\leq1.2$, all four quantities remain nearly constant, consistent
with the vortex-free, simply connected condensate observed in Fig.~\ref{fig:initial-selective}.
Beyond $\Omega=1.2$, the entropies and mutual information exhibit a
pronounced change in their behavior and increase nonlinearly with
$\Omega$. The nearly identical behavior of $S_x$ and $S_y$ reflects the
rotational symmetry of the initial state, while the simultaneous increase
of $S_{xy}$ and $I$ indicates enhanced spatial delocalization and
correlations between the two spatial degrees of freedom. Thus, the IT
measures provide a quantitative global characterization of the spatial
reorganization accompanying vortex formation, complementing the direct
identification of vortices from the density and phase profiles.

\subsection{Interaction quench dynamics}

We first investigate the response of the condensate to an interaction
quench, in which the interaction strength is suddenly increased by
$20\%$. For vortex-free states, this perturbation predominantly induces
isotropic breathing-like dynamics, while increasing the rotation
frequency leads to the gradual emergence of additional surface
excitations. We find that multicharged (giant) vortices remain robust
under such moderate interaction changes. However, for stronger quenches
($g^{\prime}\approx5g$), the outer annular density becomes unstable and
undergoes pronounced spatial splitting.

To illustrate these distinct dynamical regimes, we focus on two
representative rotation frequencies: $\Omega=0.5$, corresponding to a
vortex-free initial state, and $\Omega=2.0$, corresponding to a giant
vortex with charge $\ell=8$.

\subsubsection{{\emph{$\Omega=0.5$: Vortex-free state and breathing mode}}}

\begin{figure}[htbp]
    \centering
    \includegraphics[width=\columnwidth]{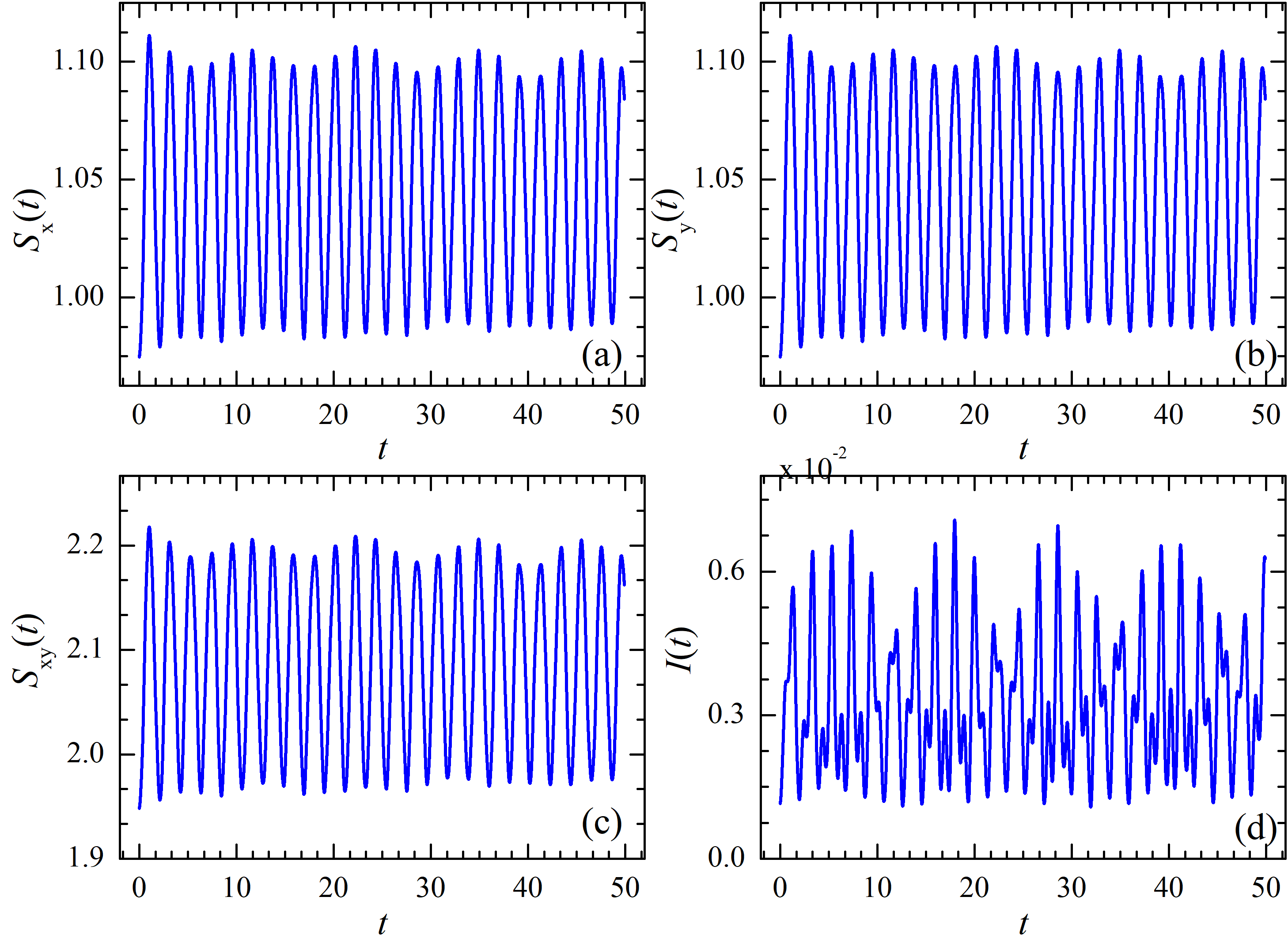}
    \caption{Time evolution of information-theoretic measures for $\Omega=0.5$ during the monopole dynamics. Panels show (a) $S_x$, (b) $S_y$, (c) $S_{xy}$, and  (d) mutual information $I$ as functions of time. 
    }
    \label{fig:mono-info-05}
\end{figure}

For $\Omega=0.5$, the condensate exhibits a pronounced monopole
oscillation following the interaction quench, with the central density
undergoing periodic expansion and contraction characteristic of a
breathing mode (not shown here). The corresponding evolution of the
IT measures is presented in Fig.~\ref{fig:mono-info-05}. All four
quantities, $S_x$, $S_y$, $S_{xy}$, and the mutual information $I$,
oscillate in a nearly periodic manner, reflecting the coherent
redistribution of density during the monopole dynamics. The nearly
identical behavior of $S_x$ and $S_y$ confirms that the isotropic
symmetry of the trap is preserved throughout the evolution, while the
small magnitude of $I$ indicates weak correlations between the $x$ and
$y$ degrees of freedom. Thus, the IT measures consistently characterize
the regular, symmetry-preserving breathing dynamics of the vortex-free
state.

The absence of significant many-body fragmentation is further confirmed
by the natural-orbital occupations shown in Appendix~A
(Fig.~\ref{fig:no-pr-mp-0.5}). The system remains predominantly
condensed, with the leading natural orbital maintaining an occupation
close to unity throughout the evolution, while the higher orbitals
exhibit only small-amplitude oscillations with negligible populations.

\begin{figure}[htbp]
    \centering
    \includegraphics[width=\columnwidth]{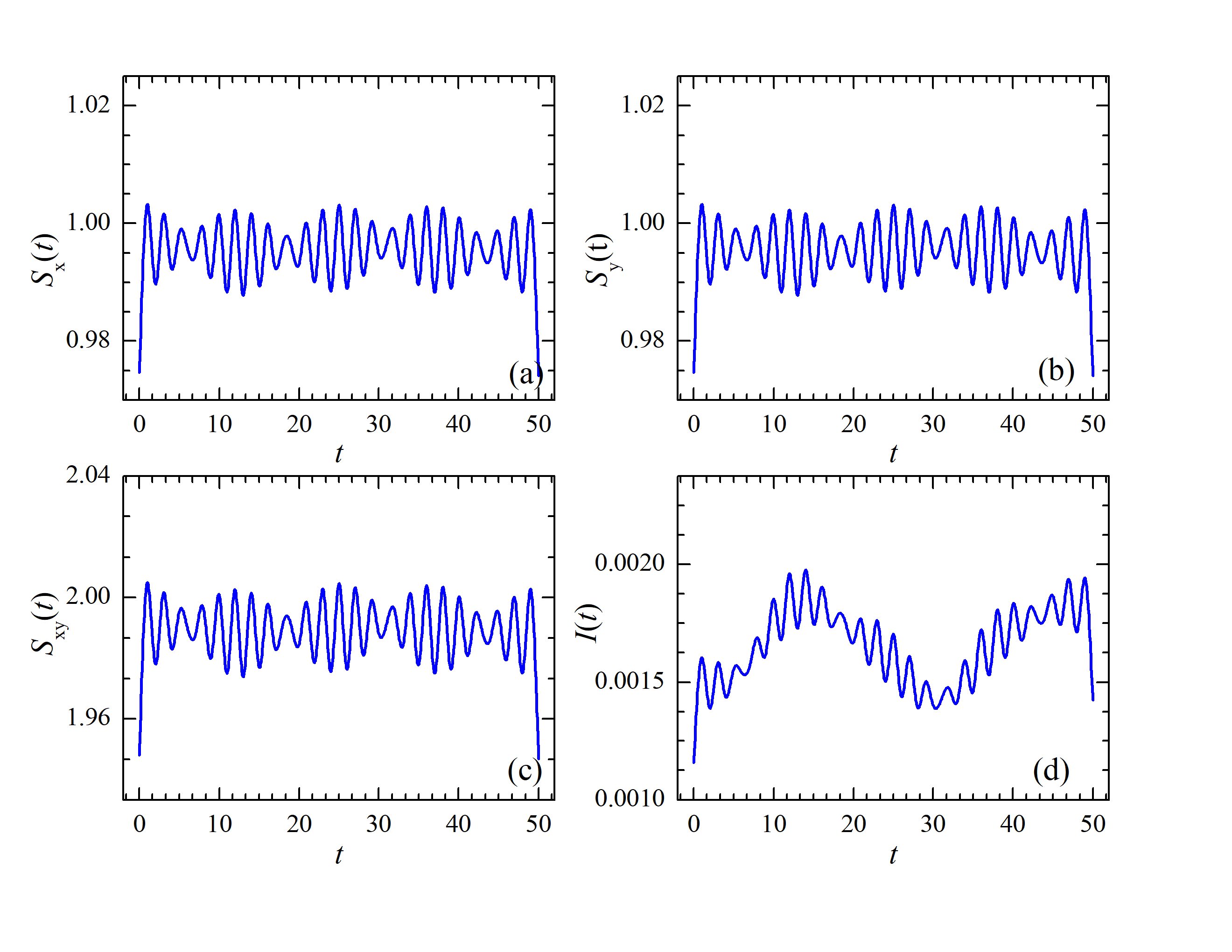}
    \caption{Time evolution of the information-theoretic measures for
the mean-field  dynamics ($M=1$) at $\Omega=0.5$ following the interaction quench. Panels show (a) $S_x$,
(b) $S_y$, (c) $S_{xy}$, and (d) the mutual information $I$
as functions of time.}
    \label{fig:GP-mono-info-05}
\end{figure}

To further examine the role of beyond-mean-field effects, we compare the information-theoretic dynamics obtained from MCTDHB and the mean-field ($M=1$) approximation following the same interaction-quench protocol. Fig.~\ref{fig:GP-mono-info-05} shows the corresponding entropy dynamics for the vortex-free initial state at \(\Omega=0.5\). While the MCTDHB results exhibit a continuous recurring periodic breathing pattern, mean-field dynamics exhibits piecewise periodic breathing. This difference indicates that the recurrent entropy oscillations observed in the correlated calculation are not captured by the mean-field description and highlight the role of many-body correlations in the post-quench evolution.

\begin{figure}[htbp]
    \centering
    \includegraphics[width=\columnwidth]{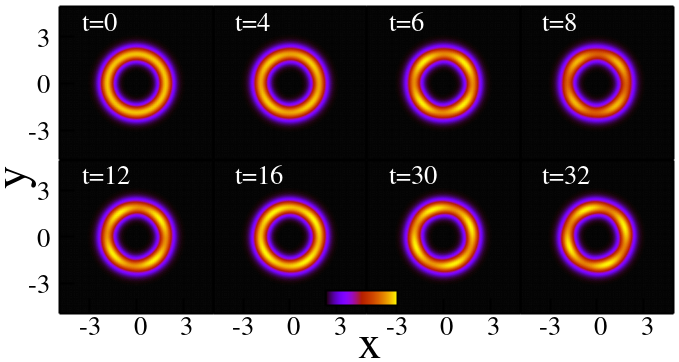}
    \caption{Time evolution of the density for the $\Omega=2.0$ condensate following a stronger interaction quench ($g^{\prime} \approx 5 g$). The central giant vortex (dark circular region) remains largely circular, while the bright outer ring develops a fourfold modulation, splitting into four partially connected high-density segments. The snapshots illustrate the combined effects of radial breathing and angular symmetry breaking at selected times.}
    \label{fig:mono-density-20}
\end{figure}

\subsubsection{{\emph{$\Omega=2.0$: Giant vortex and outer-ring splitting}}}

For $\Omega=2.0$, the condensate initially forms a giant central vortex
surrounded by an annular density distribution. Under a weak interaction
quench, the vortex remains robust and exhibits only weak breathing-like
dynamics (not shown here). In contrast, a stronger interaction quench
($g^{\prime}\approx5g$) leads to pronounced modulations of the outer
density ring. Fig.~\ref{fig:mono-density-20} shows density snapshots at
selected times following such a strong quench. The central vortex core,
visible as a dark circular region, largely preserves its shape, exhibiting
only minor deformations during the evolution. At the same time, the outer
annulus develops a clear fourfold modulation, with the density
redistributing into four high-density segments that remain partially
connected, producing a pronounced fourfold spatial deformation of the
outer ring.

\begin{figure}[htbp]
    \centering
    \includegraphics[width=\columnwidth]{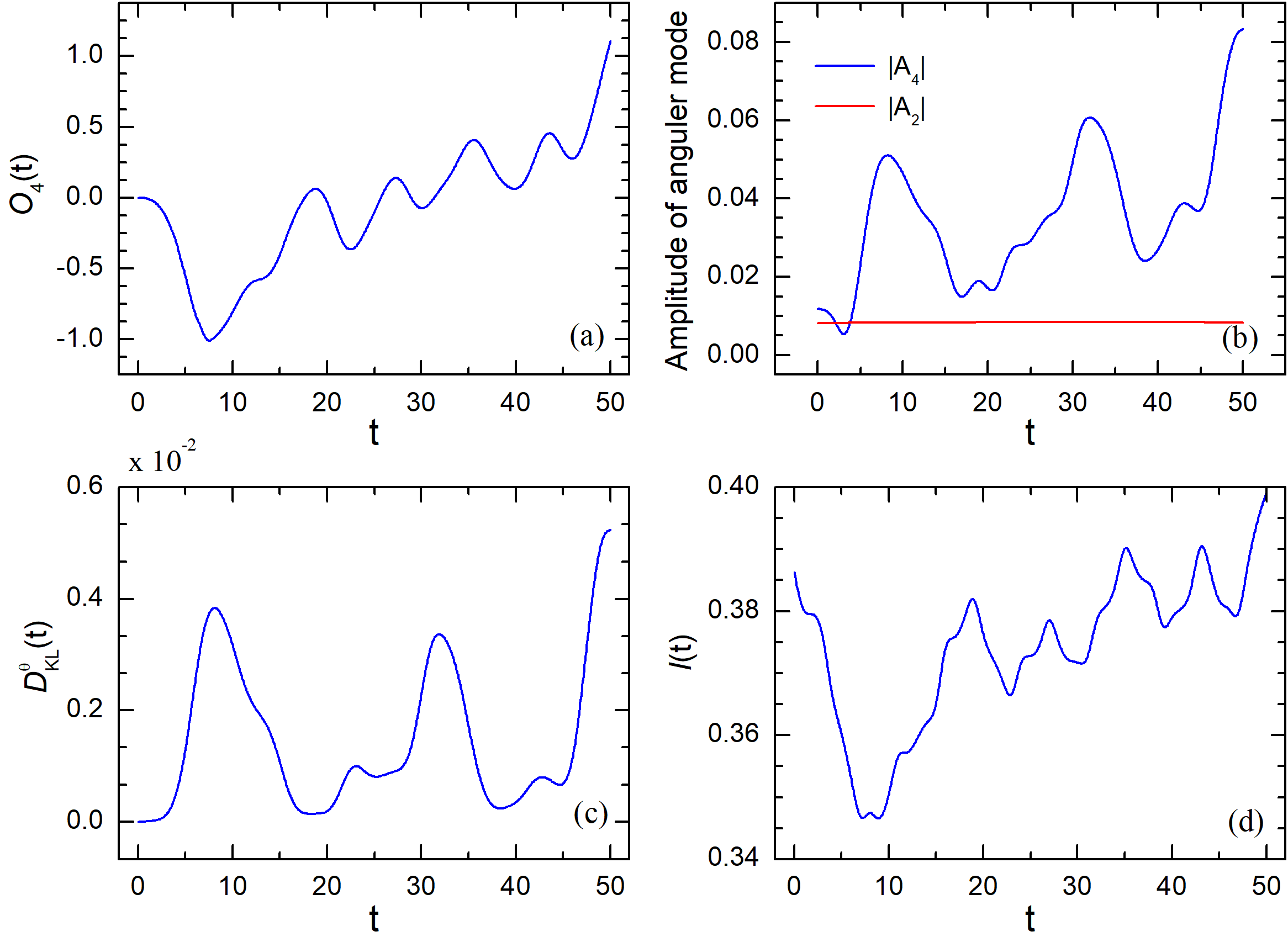}
\caption{Time evolution of angular observables along the outer ring. We follow the same quench protocol as in Fig.~\ref{fig:mono-density-20}. Observables: (a) Radially weighted fourfold operator $O_4(t)$. (b) Angular Fourier components $|A_4(t)|$ and $|A_2(t)|$. (c) Angular Kullback--Leibler divergence $D_{\mathrm{KL}}^{\theta}(t)$, evaluated over the radial interval $r_0-\Delta r\leq r\leq r_0+\Delta r$, with $r_0=1.889$ and $\Delta r=0.536$. (d) Mutual information $I(t)$.}
\label{fig:mono-info-20}
\end{figure}

To quantify the angular symmetry breaking, we monitor four complementary
observables (Fig.~\ref{fig:mono-info-20}). The radially weighted fourfold
operator $O_4(t)$ exhibits oscillations in both magnitude and sign,
reflecting changes in the orientation of the density maxima along the
ring. The angular Fourier component $A_4(t)$ shows a pronounced increase
with temporal oscillations, capturing the growth and modulation of the
fourfold angular structure, while the negligible $A_2(t)$ confirms the
absence of significant elliptic deformation.

The angular Kullback--Leibler divergence $D_{\mathrm{KL}}^{\theta}(t)$
also exhibits pronounced temporal oscillations, with a structure similar
to that of $|A_4(t)|$. This correspondence indicates that the dominant
angular redistribution is governed by the emerging fourfold component.
Thus, $D_{\mathrm{KL}}^{\theta}$ provides an IT quantification of the angular redistribution associated with the
symmetry breaking captured by $A_4$. The mutual information
$I(t)=S_x+S_y-S_{xy}$ exhibits pronounced oscillations as well, indicating
the buildup of correlations between the $x$ and $y$ degrees of freedom
during the development of the anisotropic density structure.

The correspondence among these observables highlights their
complementary roles. While $A_4$ and $O_4$ directly characterize the
fourfold density structure and its orientation, $D_{\mathrm{KL}}^{\theta}$
quantifies the corresponding redistribution of the angular probability
density. The similar temporal behavior of $D_{\mathrm{KL}}^{\theta}$ and
$|A_4|$ demonstrates that the IT measure remains sensitive to the same
symmetry-breaking dynamics identified by the dominant angular mode,
while the mutual information provides an additional global measure of
the correlations generated between the two spatial degrees of freedom.

The analysis of the natural-orbital occupations reveals significant
many-body fragmentation, as shown in Appendix~A
(Fig.~\ref{fig:no-pr-mp-2.0}). In contrast to the vortex-free regime,
multiple natural orbitals acquire appreciable occupations, indicating a
substantial redistribution of particles among the orbitals and the
development of pronounced many-body fragmentation.

The many-body fragmentation accompanying this dynamics is
independently confirmed by the natural-orbital occupations,
which show appreciable population of several orbitals.
Thus, the observed fragmentation is not inferred from the
spatial splitting of the density or from the information-theoretic
measures alone.

\begin{figure}[htbp]
    \centering
    \includegraphics[width=\columnwidth]{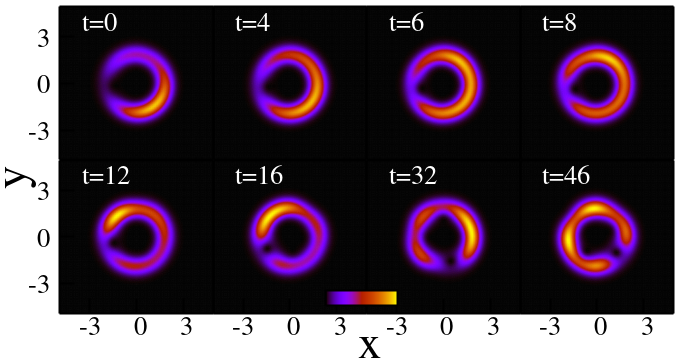}
    \caption{Time evolution of the one-body density obtained within the
mean-field approximation ($M=1$) for the $\Omega=2.0$
condensate following a stronger interaction quench
($g^{\prime}\approx 5g$). The initial state is highly
unconverged. Although an apparent splitting of the outer
density ring is visible during the subsequent evolution, this
feature may be an artifact of the unconverged initial state.}
    \label{fig:GP-mono-density-20}
\end{figure}

We also examine the mean-field dynamics for the giant-vortex initial state at \(\Omega=2.0\), as illustrated in Fig.~\ref{fig:GP-mono-density-20}. The mean-field density exhibits spatial splitting following the interaction quench. However, the corresponding mean-field initial state is not sufficiently converged, as already indicated by the comparison of the initial states. Consequently, the subsequent density evolution cannot be regarded as a reliable mean-field description of the giant-vortex dynamics. In particular, the apparent density splitting should not be interpreted as a quantitatively meaningful prediction of the instability observed in the converged MCTDHB calculation.

\begin{figure}[htbp]
    \centering
    \includegraphics[width=\columnwidth]{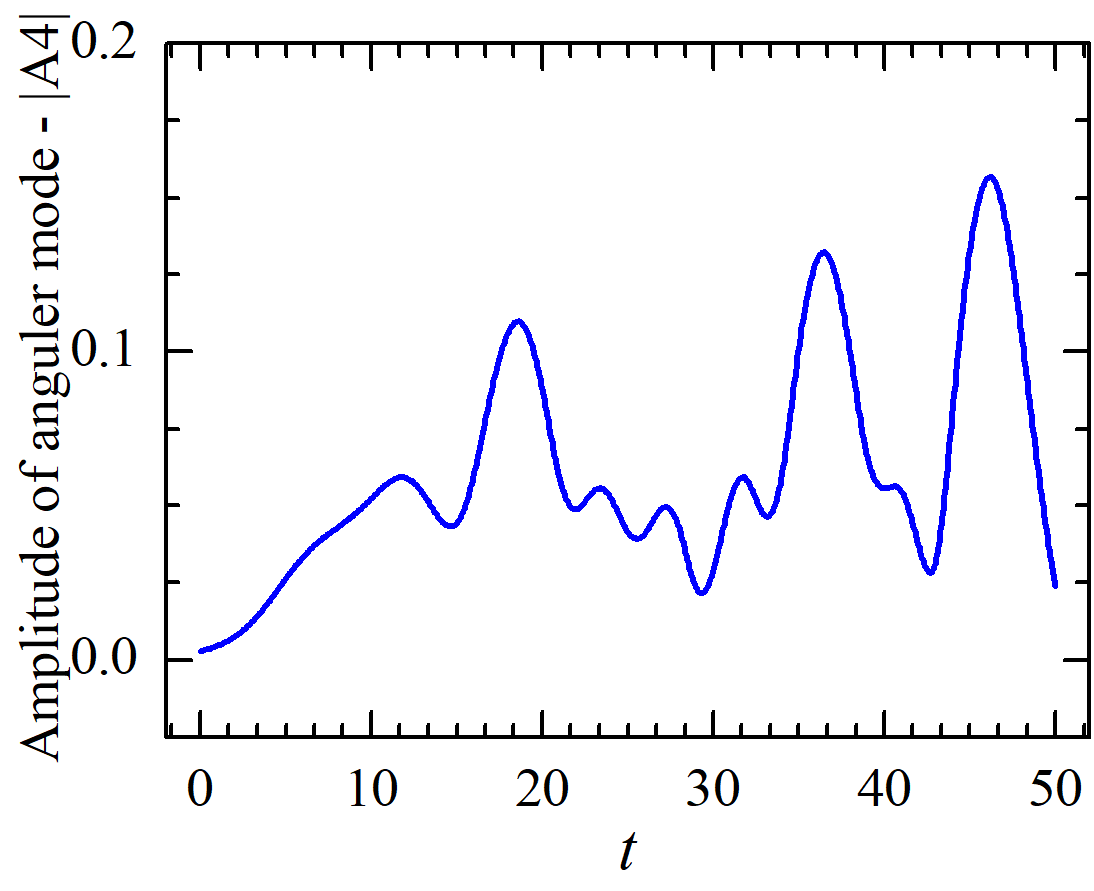}

\caption{Time evolution of the fourfold angular Fourier component
$|A_4(t)|$ obtained within the mean-field approximation
($M=1$) for $\Omega=2.0$, following the stronger interaction
quench ($g'\approx 5g$).}

    \label{fig:GP-A4}
\end{figure}

The corresponding evolution of the fourfold angular component \(A_4(t)\) is shown in Fig.~\ref{fig:GP-A4}. A finite fourfold contribution develops during the mean-field evolution, consistent with the apparent spatial splitting of the density. However, given the lack of convergence of the initial mean-field state at \(\Omega=2.0\), this response cannot be considered a reliable reproduction of the fourfold instability obtained from MCTDHB. Rather, it illustrates the sensitivity of the subsequent mean-field dynamics to the quality of the initial state. Thus, a quantitative comparison of the giant-vortex instability requires a sufficiently converged initial state, which is not achieved within the present mean-field calculation.

%

\subsection{Trap Quench Dynamics}

Next, we investigate the condensate response to a quadrupole excitation,
induced by an instantaneous reduction of the quadratic trap coefficient
along the $x$ direction, from $1.0$ to $0.8$, thereby elongating the trap
along this direction:
$V'_{\mathrm{pot}}(x,y)=\frac{1}{4}(0.8x^2+y^2)^2$. We focus on the same
two representative rotation frequencies considered for the interaction
quench, $\Omega=0.5$ and $\Omega=2.0$. For $\Omega=0.5$, corresponding to
a vortex-free initial state, the rotating condensate exhibits regular
quadrupole oscillations. In contrast, for $\Omega=2.0$, where the initial
state hosts a multicharged giant vortex, the dynamics become strongly
irregular and exhibit complex multimode behavior. This transition from
regular to irregular quadrupole dynamics is further characterized by the
mutual information $I(t,\Omega)$, whose dynamical map in the
$t$--$\Omega$ plane reveals a clear distinction between the regular
vortex-free regime and the strongly correlated dynamics of the
multicharged-vortex regime. \\

\subsubsection{\emph{$\Omega$=0.5: Vortex-free state and quadrupole mode}}

\begin{figure}[htbp]
    \centering
    \includegraphics[width=\columnwidth]{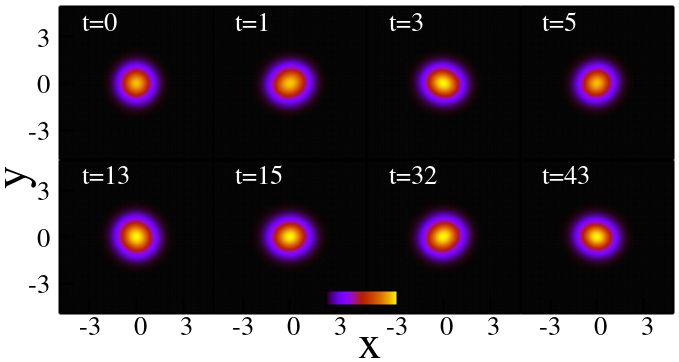}
    \caption{Density snapshots at selected times for $\Omega=0.5$ following the trap quench. The initially circular, vortex-free cloud oscillates between prolate and oblate shapes along $x$ and $y$ axes. The density remains compact with no fragmentation, and the center of mass stays near the trap origin, indicating a nearly pure quadrupole oscillation.}
    \label{fig:density-QP-omega=0.5}
\end{figure}

For $\Omega=0.5$, the initial condensate is vortex-free and nearly
circular. Following the trap quench along the $x$ direction, the density
evolves periodically, as shown in Fig.~\ref{fig:density-QP-omega=0.5}.
The initially circular cloud deforms into an ellipse oriented along the
$x$ axis, swinging to the right, then contracting along $x$ and
elongating along $y$, before returning to the left in a smooth, regular
cycle. The oscillation alternates between prolate and oblate shapes along
the two axes, while the center of mass remains essentially fixed,
consistent with a predominantly quadrupolar response. Throughout the
evolution, the density remains compact without visible spatial density
splitting, and the effects of rotation are weak, consistent with the low
rotation frequency. This behavior contrasts sharply with higher-$\Omega$
states, where the presence of a multicharged vortex can induce complex
and irregular quadrupole dynamics.

\begin{figure}[htbp]
    \centering
    \includegraphics[width=\columnwidth]{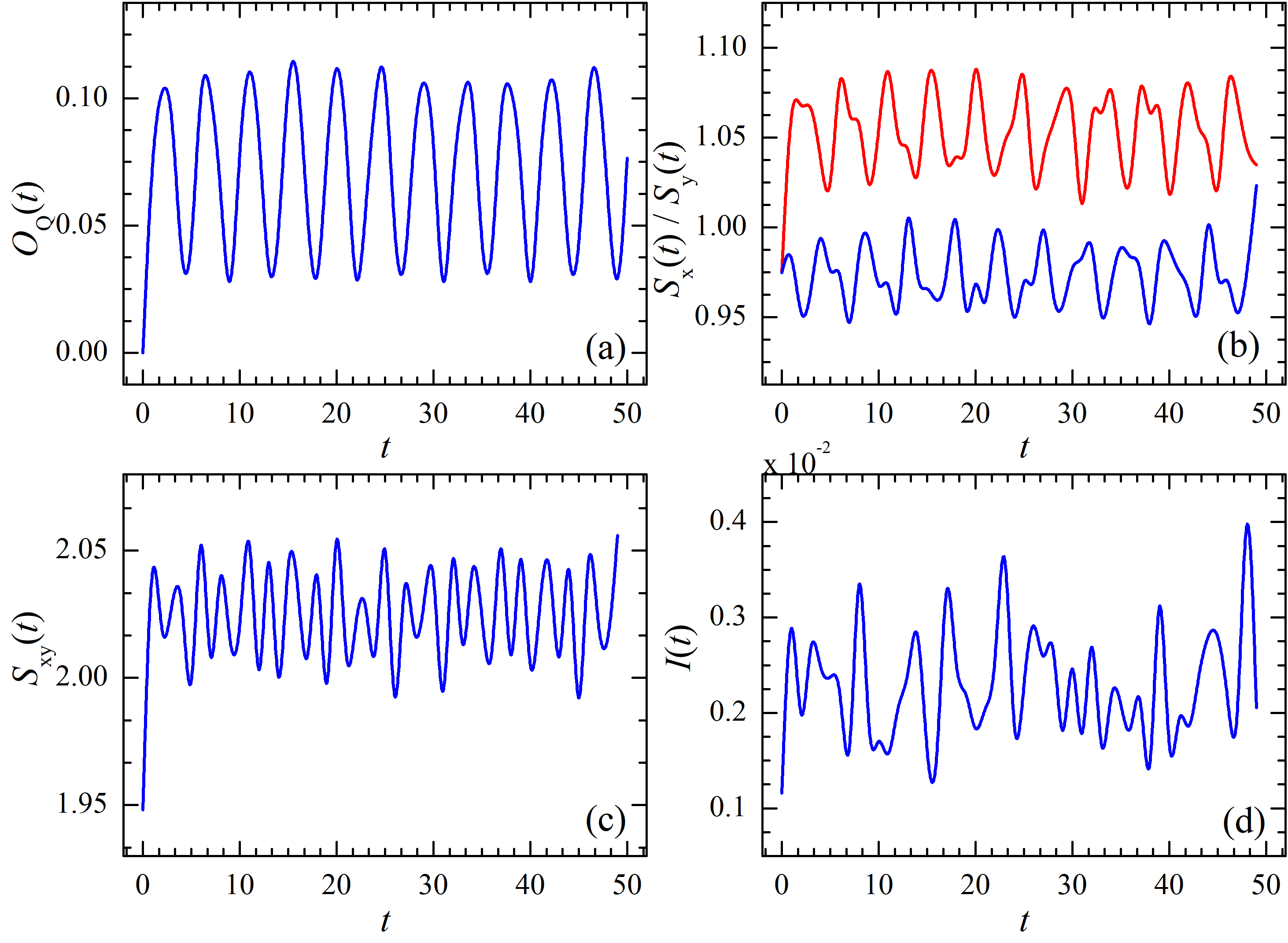}
    \caption{Time evolution of quadrupole and information-theoretic observables for $\Omega=0.5$ following the trap quench. (a) Quadrupole operator $O_Q(t)$,
(b) marginal entropies $S_x(t)$ (red curve) and  $S_y(t)$ (blue curve), (c) joint entropy $S_{xy}(t)$, and (d) mutual information $I(t)$
}
    \label{fig:quad-info-05}
\end{figure}

To further characterize the quadrupole dynamics, we analyze the time
evolution of collective and IT observables, as shown in
Fig.~\ref{fig:quad-info-05}. The quadrupole operator $O_Q(t)$ [panel (a)]
exhibits regular periodic oscillations, confirming the coherent
quadrupolar response. In panel (b), the marginal entropies $S_x(t)$ and
$S_y(t)$ display nearly identical oscillations with opposite phase,
reflecting the alternating elongation and compression along the two
spatial directions. When $S_x$ reaches a minimum, $S_y$ attains a
maximum, indicating that localization along one direction is accompanied
by delocalization along the perpendicular direction during the
quadrupole oscillation. The joint entropy $S_{xy}(t)$ [panel (c)] also
oscillates periodically, tracking the redistribution of density during
the collective motion. Finally, the mutual information $I(t)$ [panel
(d)] remains small and exhibits weak, regular oscillations, indicating
weak correlations between the $x$ and $y$ degrees of freedom. Thus, the
IT measures provide a consistent quantitative characterization of the
regular, symmetry-preserving quadrupole dynamics at $\Omega=0.5$.

For the trap quench, the mean-field dynamics at $\Omega=0.5$ exhibit
a markedly different response from the MCTDHB results.
While the correlated dynamics show an evolution between
prolate and oblate density distributions, the mean-field condensate
remains essentially stable following the quench. Neither
the one-body density nor the corresponding information-theoretic
entropy exhibits appreciable temporal variation. We note that
these mean-field results are not presented in the manuscript. This
comparison further indicates that the shape oscillations observed
in the MCTDHB dynamics are not reproduced at the mean-field
level and are associated with beyond-mean-field effects.

\subsubsection{{\emph{$\Omega=2.0$: Giant vortex and chaotic dynamics}}}
\begin{figure}[htbp]
    \centering
    \includegraphics[width=\columnwidth]{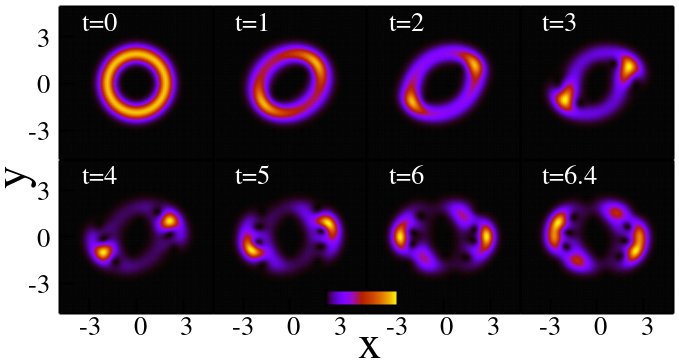}
    \caption{One-body density snapshots at selected times for the giant vortex with initial $\Omega=2.0$ following the trap quench along the $x$-axis. The initial giant vortex and the thin outer ring undergo strong spatial deformation and splitting over time, leading to irregular dynamics.}
    \label{fig:density-QP-omega=2.0}
\end{figure}

Fig.~\ref{fig:density-QP-omega=2.0} shows one-body density snapshots
at selected times following the trap quench for the rotating condensate
with initial $\Omega=2.0$. The initial state hosts a central
multicharged vortex of charge $\ell=8$, surrounded by a thin annular
density. At this high rotation frequency, the annular condensate is
highly susceptible to the imposed anisotropic perturbation. Following
the trap quench, the giant-vortex structure rapidly becomes deformed and
the central circulation breaks into smaller vortex structures.
Simultaneously, the thin outer ring develops a pronounced two-lobed
deformation during the early-time dynamics. At longer times
($t>4$), the density evolution becomes increasingly irregular, with
multiple azimuthally localized structures developing around the central
depleted region. The original annular structure does not recover, and
the vortex structures remain spatially dispersed over the timescale
considered. No clear revival of the initial configuration is observed,
indicating persistent irregular dynamics characteristic of the
chaotic quadrupole regime.

\begin{figure}[htbp]
    \centering
    \includegraphics[width=\columnwidth]{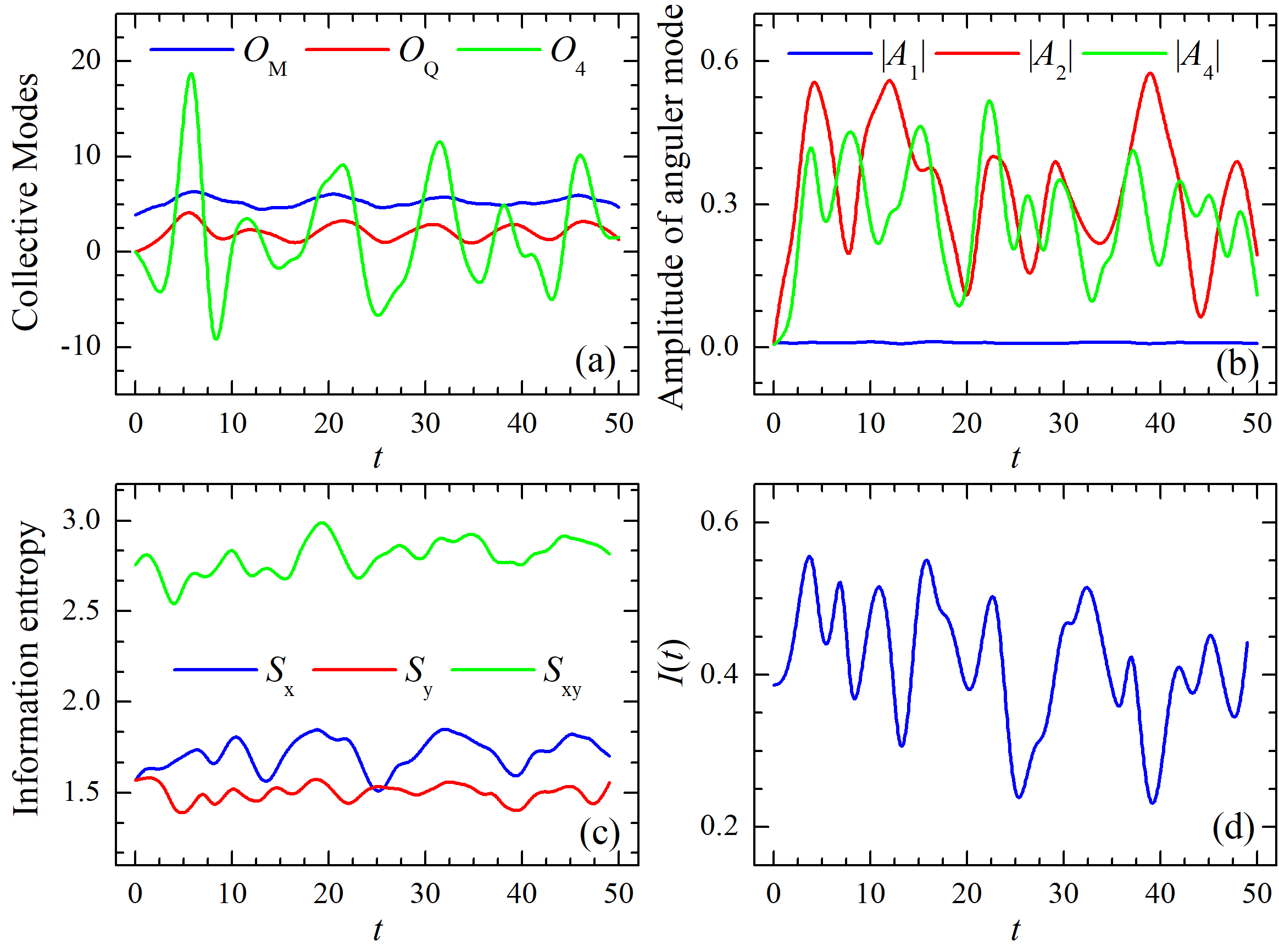}
    \caption{Time evolution of collective, angular, and IT observables for the strongly driven giant-vortex dynamics at $\Omega=2.0$. (a) Monopole, quadrupole, and fourfold operators, $O_{\mathrm M}(t)$, $O_{\mathrm Q}(t)$, and $O_4(t)$, respectively. (b) Angular Fourier components $|A_1(t)|$, $|A_2(t)|$, and $|A_4(t)|$. (c) Marginal and joint spatial entropies, $S_x(t)$, $S_y(t)$, and $S_{xy}(t)$. (d) Mutual information $I(t)$. The pronounced oscillations of the collective and angular observables, together with the fluctuations of the IT measures, characterize the multimode and strongly correlated nature of the vortex dynamics.}

    \label{fig:chaotic-observables}
\end{figure}

To quantify the dynamics at $\Omega=2.0$, we examine the time evolution
of collective, angular, and IT observables, as shown in
Fig.~\ref{fig:chaotic-observables}. Panel (a) shows the monopole,
quadrupole, and fourfold operators, $O_{\mathrm M}(t)$, $O_{\mathrm Q}(t)$,
and $O_4(t)$, respectively. All three observables exhibit pronounced
oscillations, with the fourfold operator $O_4(t)$ showing particularly
strong temporal variations, demonstrating that the fourfold deformation
remains dynamically active throughout the evolution. The simultaneous
excitation of monopole, quadrupole, and fourfold modes indicates strongly
multimode dynamics rather than the dominance of a single collective mode.
Panel (b) displays the angular Fourier components $A_1(t)$, $A_2(t)$, and
$A_4(t)$. The dipole component $A_1(t)$ remains negligible, ruling out
significant center-of-mass motion, whereas $A_2(t)$ and $A_4(t)$ exhibit
appreciable amplitudes, demonstrating the coexistence of quadrupolar and
fourfold angular structures.

The corresponding IT measures are shown in panels (c) and (d). The
marginal entropies $S_x(t)$ and $S_y(t)$ and the joint entropy
$S_{xy}(t)$ exhibit pronounced temporal fluctuations, with $S_x(t)$
showing a larger amplitude than $S_y(t)$, consistent with the stronger
density redistribution along the $x$ direction induced by the trap
quench. The mutual information $I(t)$ also displays strong temporal
fluctuations, indicating substantial correlations between the two
spatial degrees of freedom. Thus, the IT measures provide a quantitative
global characterization of the spatial redistribution and correlations
associated with the complex multimode vortex dynamics, complementing
the mode-resolved information provided by the collective and angular
observables.

However, the system remains predominantly condensed despite the
emergence of irregular vortex-splitting dynamics, as shown in
Appendix~A (Fig.~\ref{fig:no-pr-quad}). Although the one-body density undergoes pronounced spatial
deformation and splitting, the natural-orbital occupations
remain dominated by the leading orbital. This demonstrates
that the observed density restructuring does not necessarily
imply strong many-body fragmentation.

Instead
originates primarily from the intrinsic dynamical instability of the
multicharged vortex. The quadrupolar deformation explicitly breaks
rotational symmetry and couples to the angular modes of the condensate.
Such symmetry-breaking perturbations can destabilize the giant-vortex
configuration and become strongly amplified once seeded. In the present
case, the imposed anisotropy is therefore amplified through the coupled
angular dynamics, leading to the growth and interference of multiple
angular modes. This produces the observed breakup of the annular density
into irregular, time-dependent spatial structures, while the natural-
orbital occupations remain dominated by the leading orbital. The
resulting complex and effectively chaotic dynamics can therefore emerge
without pronounced many-body fragmentation, highlighting the distinction
between spatial vortex splitting and fragmentation of the many-body
state.

\begin{figure}[htbp]
    \centering
    \includegraphics[width=\columnwidth]{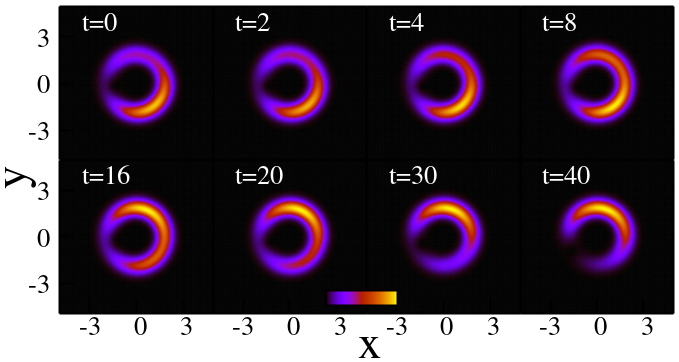}
    \caption{Mean-field ($M=1$) one-body density snapshots at selected times for the giant-vortex state with initial $\Omega=2.0$, following a trap quench along the $x$-axis. }
    \label{fig:GP-density-QP-omega=2.0}
\end{figure}

For the trap quench at $\Omega=2.0$, we also examine the
corresponding mean-field density evolution, shown in Fig.~\ref{fig:GP-density-QP-omega=2.0}.
The initial mean-field giant-vortex state is already not sufficiently
converged, as discussed above. Nevertheless, we include its
subsequent evolution for comparison with the MCTDHB results.
Unlike the MCTDHB dynamics, which exhibit pronounced spatial
splitting and deformation of the giant-vortex density, the
mean-field density does not show a comparable splitting into multiple
spatial lobes. However, because of the lack of convergence
of the initial mean-field state, this comparison cannot establish
whether the absence of such splitting is an intrinsic
limitation of the mean-field description. Rather, it
illustrates the difficulty of obtaining a reliable mean-field
description of the giant-vortex dynamics at this rotation
frequency.

\subsubsection{\emph{Dynamical Phase Diagram of Trap-Quench Dynamics}}

To explore the dynamics induced by the trap quench over the full rotation
range $0\leq\Omega\leq2.0$, we calculate the time evolution of the IT
measures $S_x(t)$, $S_y(t)$, $S_{xy}(t)$, and $I(t)$ for each initial
state. Since these quantities exhibit broadly similar trends, we focus
primarily on the mutual information $I(t,\Omega)$, which provides a
compact measure of the correlations between the two spatial degrees of
freedom.

\begin{figure}[htbp]
    \centering
    \includegraphics[width=\columnwidth]{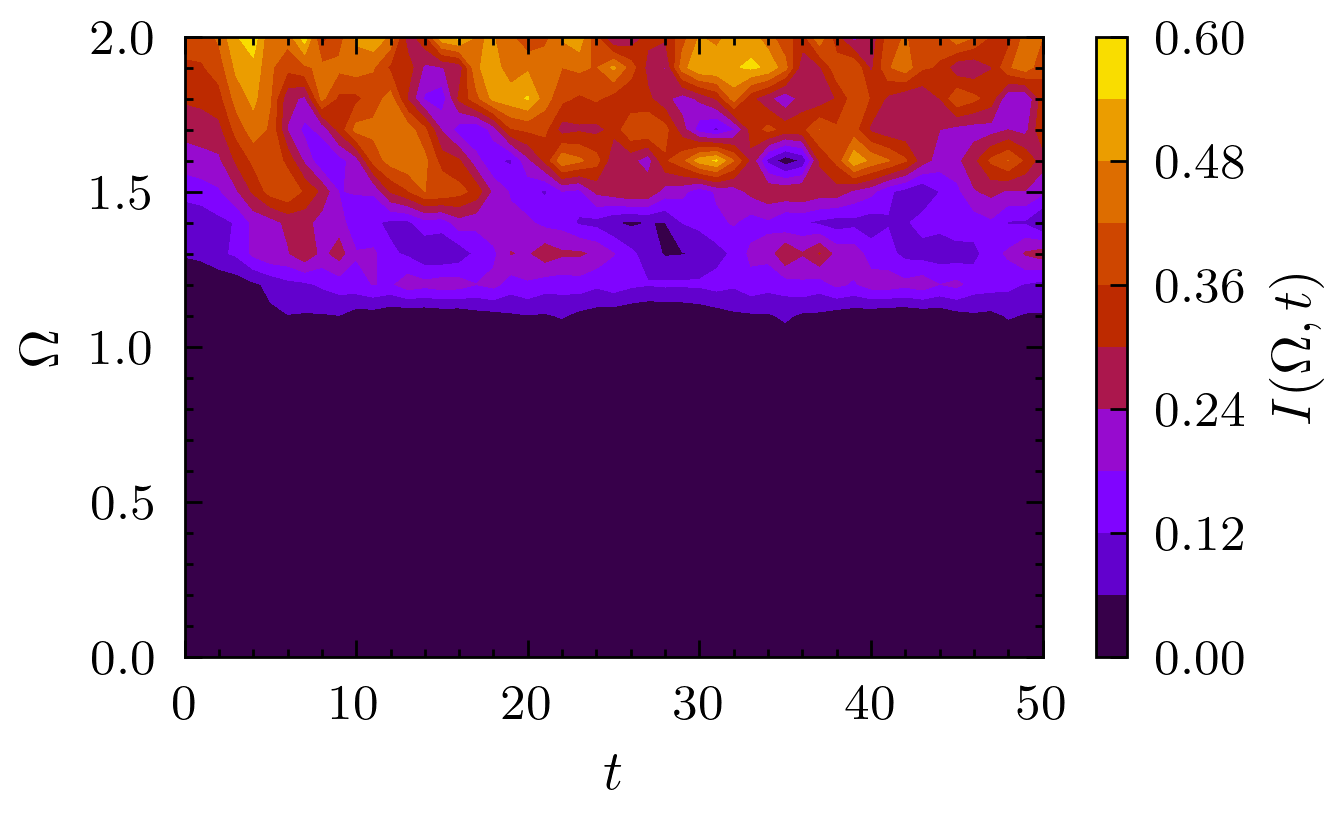}
    \caption{3D contour plot of mutual information $I(t, \Omega)$ following the trap quench. Time $t$ is along the horizontal axis, rotation frequency $\Omega$
along the vertical axis, and color represents the magnitude of $I$. The diagram highlights the transition from regular quadrupole oscillations at low $\Omega$
to chaotic dynamics at higher $\Omega$.
}
    \label{fig:phase-diagram}
\end{figure}

Using $I(t,\Omega)$, we construct the dynamical map shown in
Fig.~\ref{fig:phase-diagram}, with time $t$ on the horizontal axis,
rotation frequency $\Omega$ on the vertical axis, and the magnitude of
$I$ represented by the color scale. The map reveals three qualitatively
distinct dynamical regimes. At low rotation frequencies, corresponding
to vortex-free initial states, $I$ remains close to zero throughout the
evolution, consistent with weak correlations and regular collective
motion. At intermediate $\Omega$, $I$ develops appreciable values,
reaching up to $\sim0.24$, together with pronounced temporal variations,
indicating enhanced correlations and increasingly complex dynamics. At
higher rotation frequencies, a second regime emerges in which $I$
reaches substantially larger values, up to $\sim0.6$, and exhibits
strong temporal fluctuations. This behavior signals a pronounced buildup
of correlations associated with the complex dynamics of the
multicharged-vortex states.

Overall, the dynamical map of $I(t,\Omega)$ provides a compact global
characterization of the evolution from weakly correlated, regular
dynamics to strongly correlated and highly fluctuating dynamics as the
rotation frequency is increased. The distinct regimes identified in the
map are consistent with the changes in vortex structure and collective
mode content observed from the conventional dynamical observables.

The corresponding dynamical maps of $S_x(t)$, $S_y(t)$, and
$S_{xy}(t)$ are presented in Appendix~C. Although these measures display
similar overall trends, they provide less pronounced separation of the
intermediate dynamical regime than the mutual information.

\section{Conclusions}
\label{sec4}
In this work, we have investigated the nonequilibrium dynamics of a
two-dimensional rotating Bose--Einstein condensate following interaction
and trap quenches, using a fully correlated many-body framework. By
considering initial states ranging from vortex-free configurations to
multicharged giant vortices, we have examined how the excitation protocol
and rotation frequency govern the resulting collective and vortex
dynamics. Interaction quenches predominantly induce regular monopole-like
breathing dynamics in vortex-free states, whereas stronger quenches of
multicharged vortices drive azimuthal instabilities and pronounced
splitting of the outer density ring. Trap quenches produce regular
quadrupolar oscillations in vortex-free states, while giant vortices
exhibit strongly irregular dynamics involving the simultaneous excitation
of multiple collective and angular modes.

A central objective of this work is to establish the role of IT measures
as quantitative diagnostics complementary to conventional vortex
observables. Density and phase profiles directly characterize vortex
structures, while collective-mode operators and angular Fourier
components resolve specific excitation and symmetry channels. Natural
orbital occupations quantify many-body fragmentation. In contrast, the
marginal and joint Shannon entropies, $S_x$, $S_y$, and $S_{xy}$ quantify
global spatial redistribution, while the mutual information $I$ measures
correlations between the two spatial degrees of freedom. The
angular-resolved KL divergence $D_{\mathrm{KL}}^\theta$ provides an
additional quantitative measure of angular redistribution associated
with symmetry breaking. The comparison with density-based diagnostics
presented in Appendix~F shows that these IT measures provide complementary
information on the global spatial and correlation structure, particularly
when local density measures become strongly suppressed and insensitive.

The combined analysis also reveals an important physical distinction
between spatial density splitting and many-body fragmentation. The
irregular splitting dynamics of the giant vortex following the trap
quench occur while the system remains predominantly condensed, as
indicated by the natural-orbital occupations. Thus, complex and
effectively chaotic vortex dynamics do not necessarily require
substantial many-body fragmentation. Instead, the observed complexity can
arise from the amplification and coupling of multiple angular and
collective modes within the dominant orbital.

The IT measures further provide a quantitative characterization of the
crossover from regular to irregular dynamics. In particular, the mutual
information $I(t)$ distinguishes the weakly correlated regular regime
from the increasingly correlated intermediate and high-rotation regimes,
and its dependence on both time and rotation frequency provides a global
dynamical characterization across the parameter range. The
angular-resolved KL divergence complements this analysis by tracking the
temporal redistribution of angular probability associated with the
dominant symmetry-breaking modes. Together with the density,
mode-resolved, and natural-orbital observables, these measures allow
spatial vortex reorganization, collective-mode excitation, many-body
fragmentation, and spatial correlations to be distinguished within a
common quantitative framework.

Several questions remain open, including the long-time evolution of the
irregular vortex dynamics, the possible emergence of prethermal or
thermalized states, and the effects of finite temperature and
three-dimensional degrees of freedom. Extending the present framework to
these regimes may provide further insight into the emergence and
evolution of complexity in rotating quantum fluids.

\section*{Acknowledgments} 
L. A. Machado acknowledges support from the São Paulo Research Foundation (FAPESP) under Grant Nos. 2025/12991-9 and 2024/20641-5.  B.~ Chakrabarti acknowledges support by São Paulo Research Foundation (FAPESP) under the grants 2013/07276-1. N.~D.~Chavda acknowledges financial support from University Research
Project Grant~No.DR/Dir./26-27/17/Sr~No-5. 


\appendix

\section{Time evolution of natural orbital occupations}

We examine the time evolution of the natural orbital occupations to
distinguish spatial density splitting from dynamical many-body
fragmentation. Representative cases are considered for both interaction
and trap quenches, covering regimes in which the condensate remains
predominantly condensed and regimes where substantial population
redistribution among natural orbitals occurs.

\begin{figure}[htbp]
    \centering
    \includegraphics[width=\columnwidth]{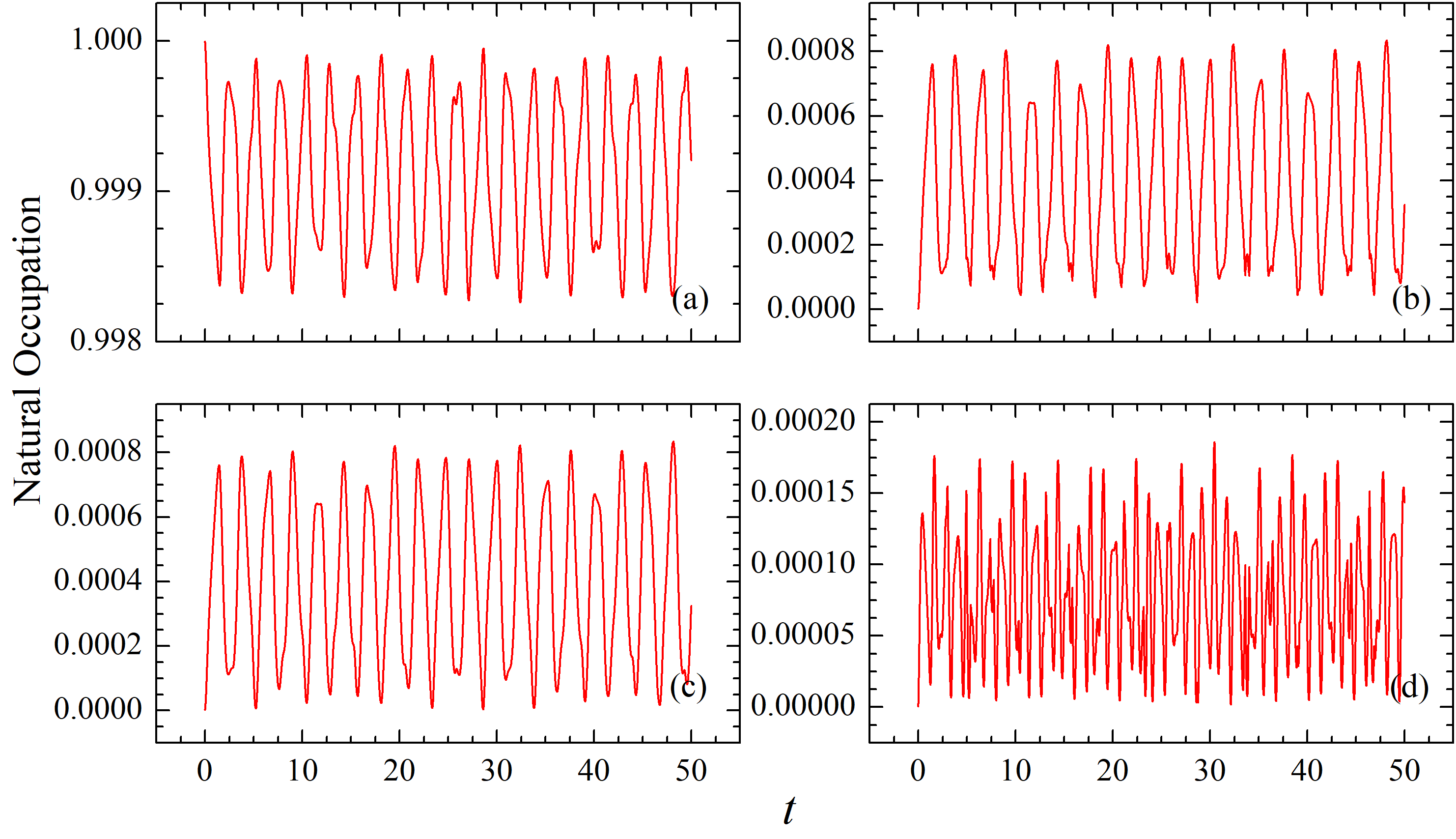}
    \caption{Time evolution of the first four natural orbital occupations following the interaction quench for the vortex-free initial state at $\Omega = 0.5$. Panels (a)--(d) show the occupations $n_1(t)$, $n_2(t)$, $n_3(t)$, and $n_4(t)$, respectively, obtained from an $M=6$ orbital description.}
    \label{fig:no-pr-mp-0.5}
\end{figure}

For the vortex-free initial state at $\Omega=0.5$ under an interaction quench, the natural-orbital occupations are shown in Fig.~\ref{fig:no-pr-mp-0.5}. The calculation uses $M=6$ orbitals, with the occupations of the first four natural orbitals displayed. The system
remains predominantly condensed throughout the dynamics, with the
leading orbital maintaining an occupation of approximately $0.99$ and
only weak temporal oscillations. The higher orbitals remain negligibly
populated. Thus, despite the excitation induced by the quench, there is
no appreciable dynamical many-body fragmentation, and the evolution
remains well described by a single dominant orbital.

\begin{figure}[htbp]
    \centering
    \includegraphics[width=\columnwidth]{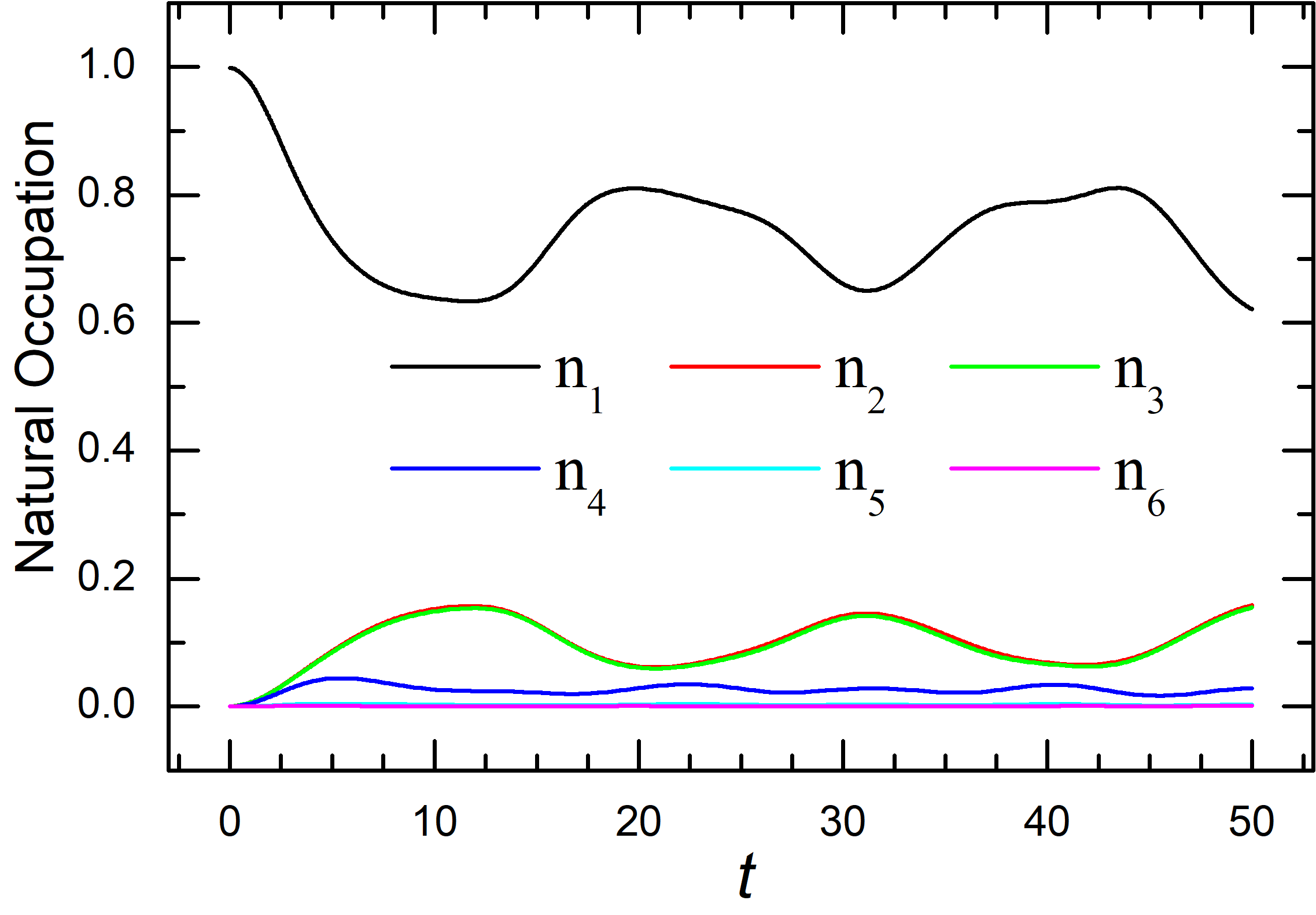}
    \caption{Time evolution of the natural orbital occupations following the strong interaction quench for the initial state at $\Omega = 2.0$. All six occupations $n_1(t)$--$n_6(t)$, obtained from an $M=6$ orbital description, are shown together in a single panel.}
    \label{fig:no-pr-mp-2.0}
\end{figure}

For $\Omega=2.0$, corresponding to an initial multicharged (giant)
vortex, the strong interaction quench produces a markedly different
response (Fig.~\ref{fig:no-pr-mp-2.0}). The leading orbital loses a
substantial fraction of its occupation, while several higher orbitals
acquire appreciable populations with pronounced temporal variations.
This redistribution demonstrates the development of significant
dynamical many-body fragmentation, accompanying the symmetry-breaking
splitting of the annular density.

\begin{figure}[htbp]
    \centering
    \includegraphics[width=\columnwidth]{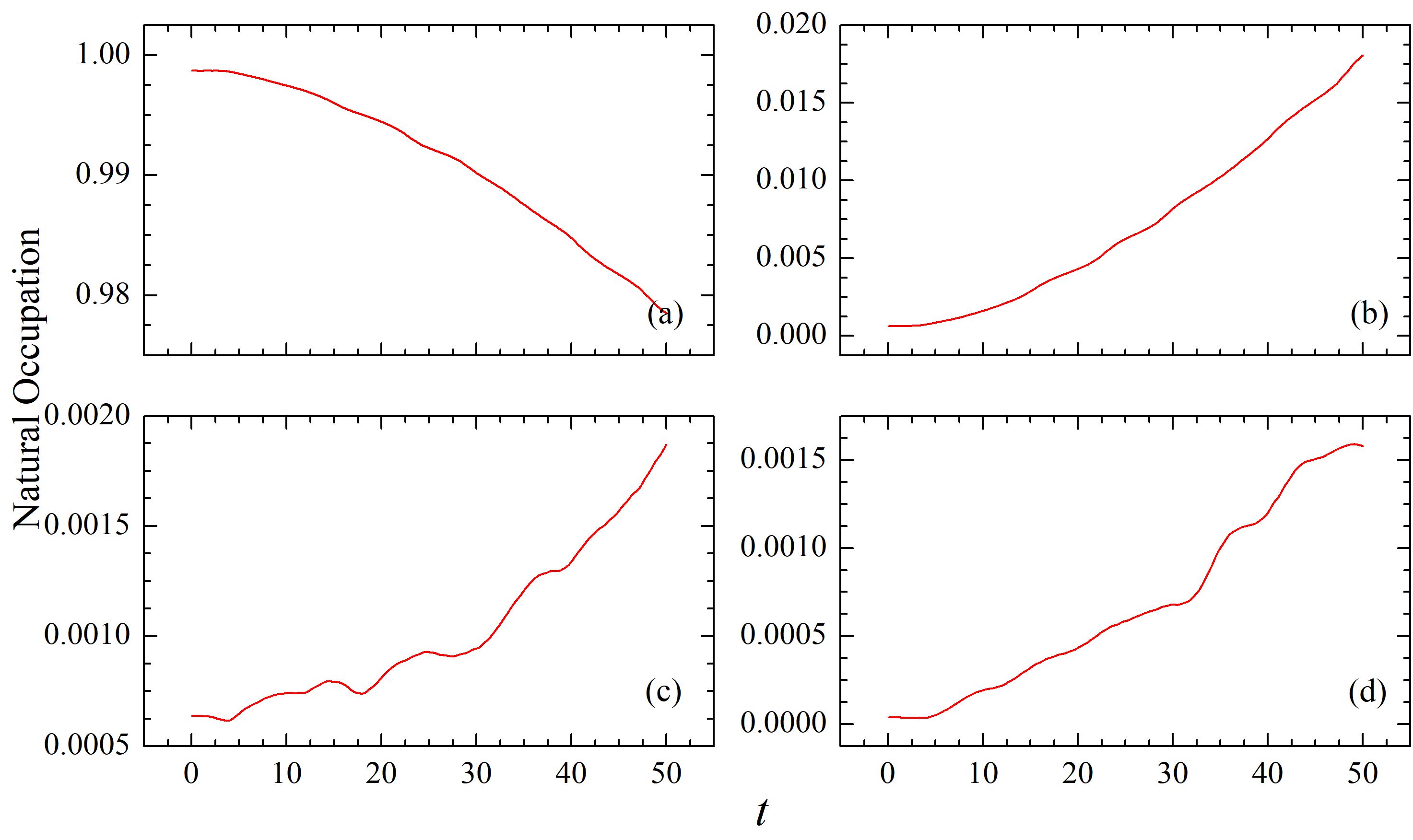}
    \caption{Time evolution of the first four natural orbital occupations following the trap quench for the initial state at $\Omega = 2.0$. Panels (a)--(d) show the occupations $n_1(t)$, $n_2(t)$, $n_3(t)$, and $n_4(t)$, respectively, obtained from an $M=6$ orbital description.}
    \label{fig:no-pr-quad}
\end{figure}

In contrast, the trap-quench dynamics at $\Omega=2.0$ do not produce
substantial many-body fragmentation (Fig.~\ref{fig:no-pr-quad}). The
leading natural orbital remains predominantly occupied, while the
higher orbitals retain only small populations. Hence, the pronounced
vortex-splitting and irregular density dynamics observed after the trap
quench cannot be attributed to significant population transfer among
natural orbitals. Instead, the instability develops predominantly within
the dominant orbital through the excitation and coupling of
symmetry-breaking angular modes. This distinction demonstrates that
spatial density splitting and many-body fragmentation represent
different aspects of the nonequilibrium dynamics.

\section{Trap Quench Dynamics at Intermediate Rotation Frequencies}

To further characterize the evolution from regular quadrupole dynamics to
the strongly irregular regime, we examine the condensate response at the
intermediate rotation frequencies $\Omega=1.3$ and $\Omega=1.5$ following
the trap quench. While the main text focuses on $\Omega=0.5$ and
$\Omega=2.0$ to illustrate the limiting regular and irregular regimes,
these intermediate cases provide additional insight into the progressive
development of angular deformations and mode coupling. In this regime,
the density develops increasingly pronounced anisotropic structures,
while partial revivals indicate that the dynamics still retain coherent
collective-mode character.

\begin{figure}[htbp]
    \centering
    \includegraphics[width=\columnwidth]{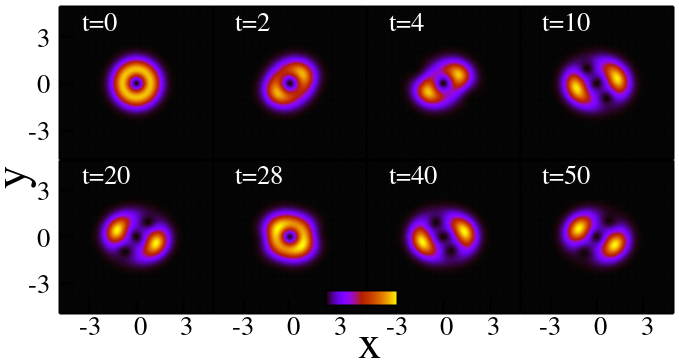}
    \caption{One-body density snapshots of the condensate at intermediate rotation frequency $\Omega=1.3$ following the trap quench, shown at different times. The density develops a two-lobed structure at intermediate times while maintaining overall coherence, illustrating the gradual transition from regular quadrupole oscillations to more complex dynamics.}
    \label{fig:density-QP-omega=1.3}
\end{figure}

For $\Omega=1.3$, the one-body density evolution is shown in
Fig.~\ref{fig:density-QP-omega=1.3}. At early times ($t=2,4$), the
condensate develops an elliptic deformation while retaining the central
density depletion. At intermediate times ($t=10,20$), the annular
density evolves into a two-lobed structure, although the lobes remain
connected. At longer times, a partial revival of the annular density is
observed, indicating that the dynamics remain dominated by a limited
number of collective modes rather than exhibiting the strongly irregular
behavior found at $\Omega=2.0$.

\begin{figure}[htbp]
    \centering
    \includegraphics[width=\columnwidth]{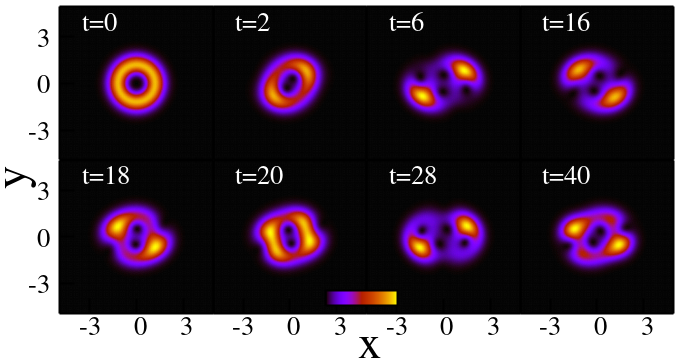}
    \caption{One-body density snapshots of the condensate at intermediate rotation frequency $\Omega=1.5$ following the trap quench, shown at different times. The initial fourfold-charged vortex splits into four singly charged vortices, and the density develops multiple lobes with azimuthal localization, illustrating the interplay between quadrupole oscillations and vortex dynamics.}
    \label{fig:density-QP-omega=1.5}
\end{figure}

For $\Omega=1.5$, the corresponding density evolution is shown in
Fig.~\ref{fig:density-QP-omega=1.5}. The initial state contains a
multicharged vortex with charge $\ell=4$, which becomes unstable under
the imposed anisotropic perturbation. At early times ($t=6$), the
central vortex splits into four singly charged vortices, accompanied by
elliptic deformation of the surrounding density. The density subsequently
develops pronounced two-lobed structures, together with an overall
rotation of the density and embedded vortices. This behavior indicates
strong coupling between the quadrupolar excitation and the angular
degrees of freedom. At later times ($t=16,18$), partial reformation of
the annular density and vortex merging are observed, although these
revivals remain transient and are followed by renewed deformation of the
density distribution.

These intermediate cases demonstrate that the response to the trap
quench evolves continuously with increasing rotation frequency: angular
deformations and vortex instabilities become progressively more
pronounced, while the dynamics gradually lose the simple periodic
character of the low-rotation regime. The strongly irregular multimode
dynamics observed at $\Omega=2.0$ therefore emerge from this progressive
increase in angular-mode coupling rather than from an abrupt change
between two qualitatively disconnected regimes.

\section{Dynamical Phase Diagrams from $S_x(t)$, $S_y(t)$, and $S_{xy}(t)$}

\begin{figure}[htbp]
    \centering
    
    \includegraphics[width=\columnwidth]{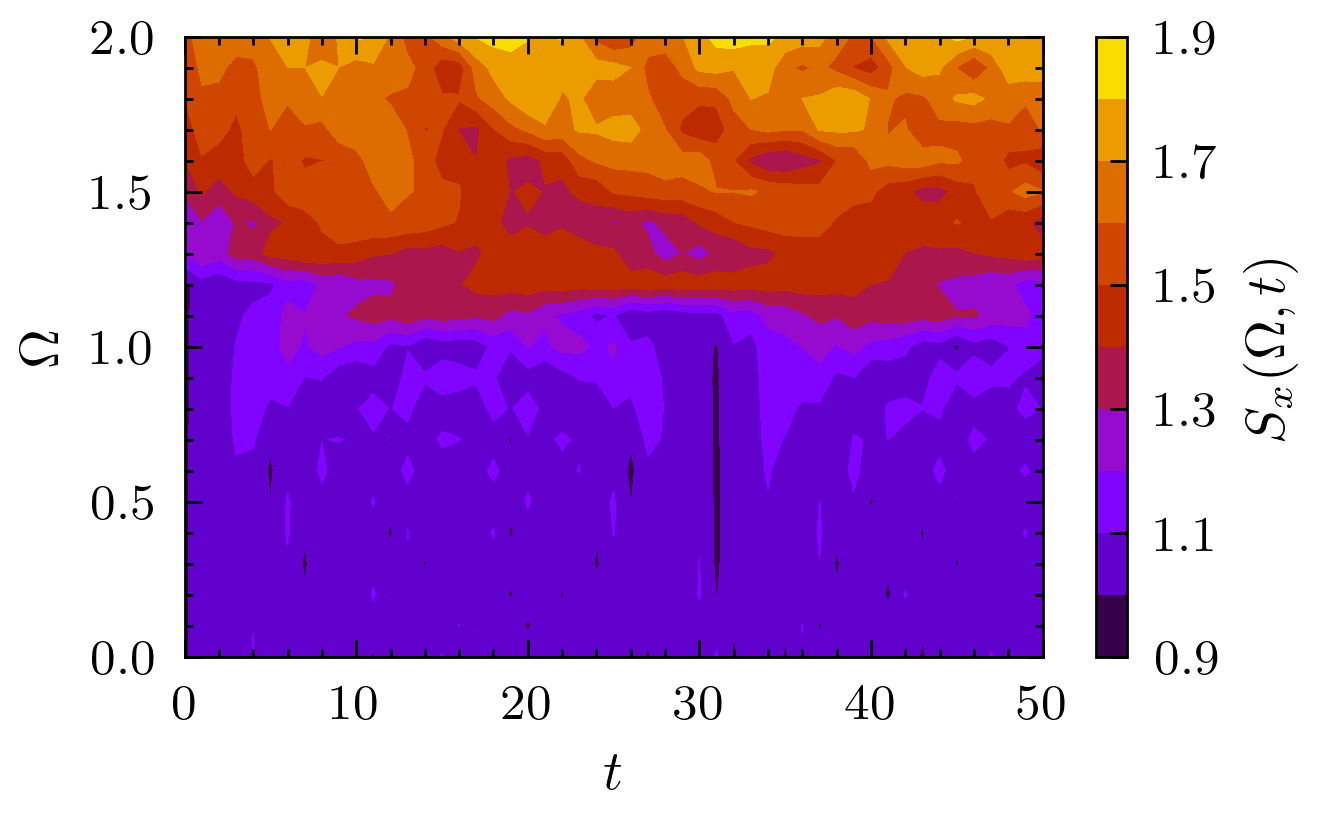}\\
    \includegraphics[width=\columnwidth]{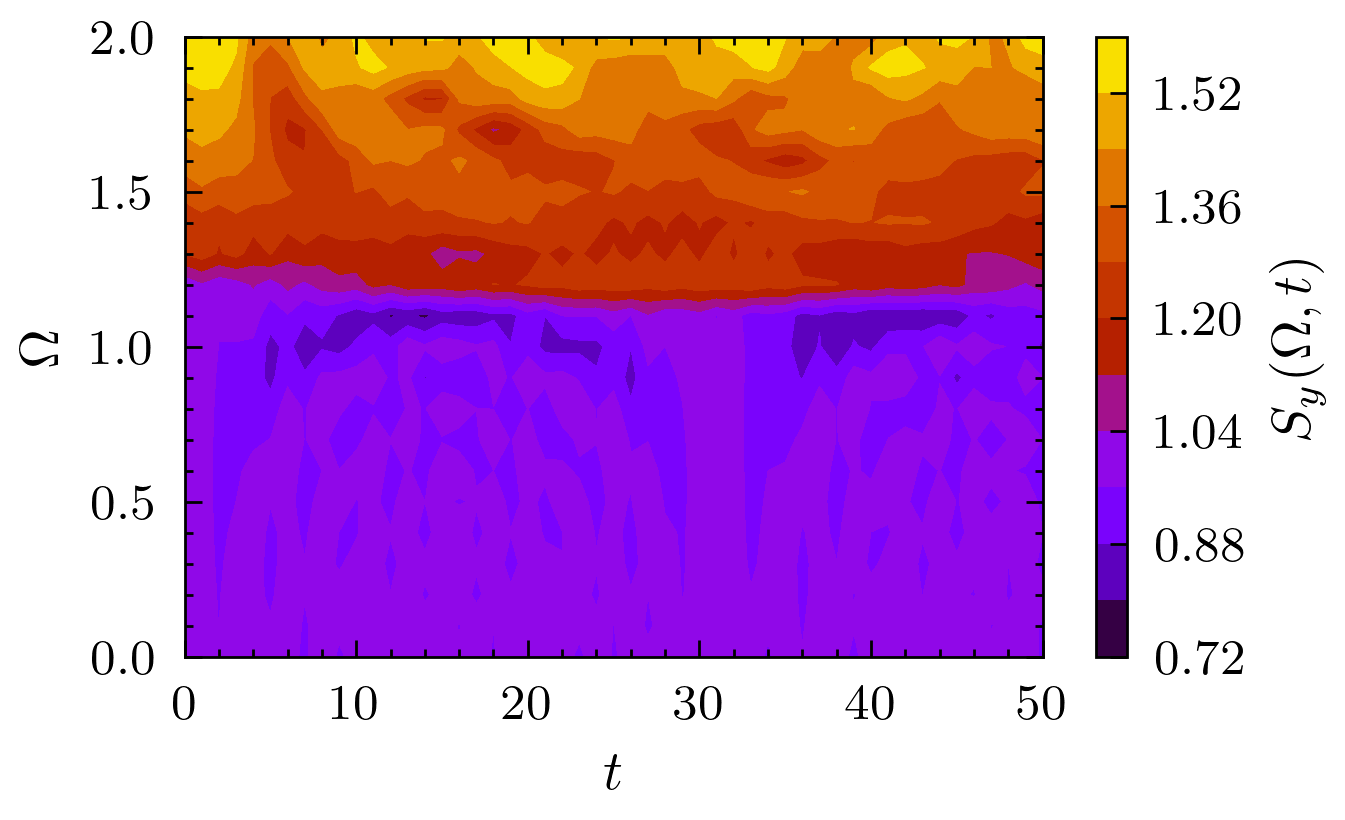}\\
    \includegraphics[width=\columnwidth]{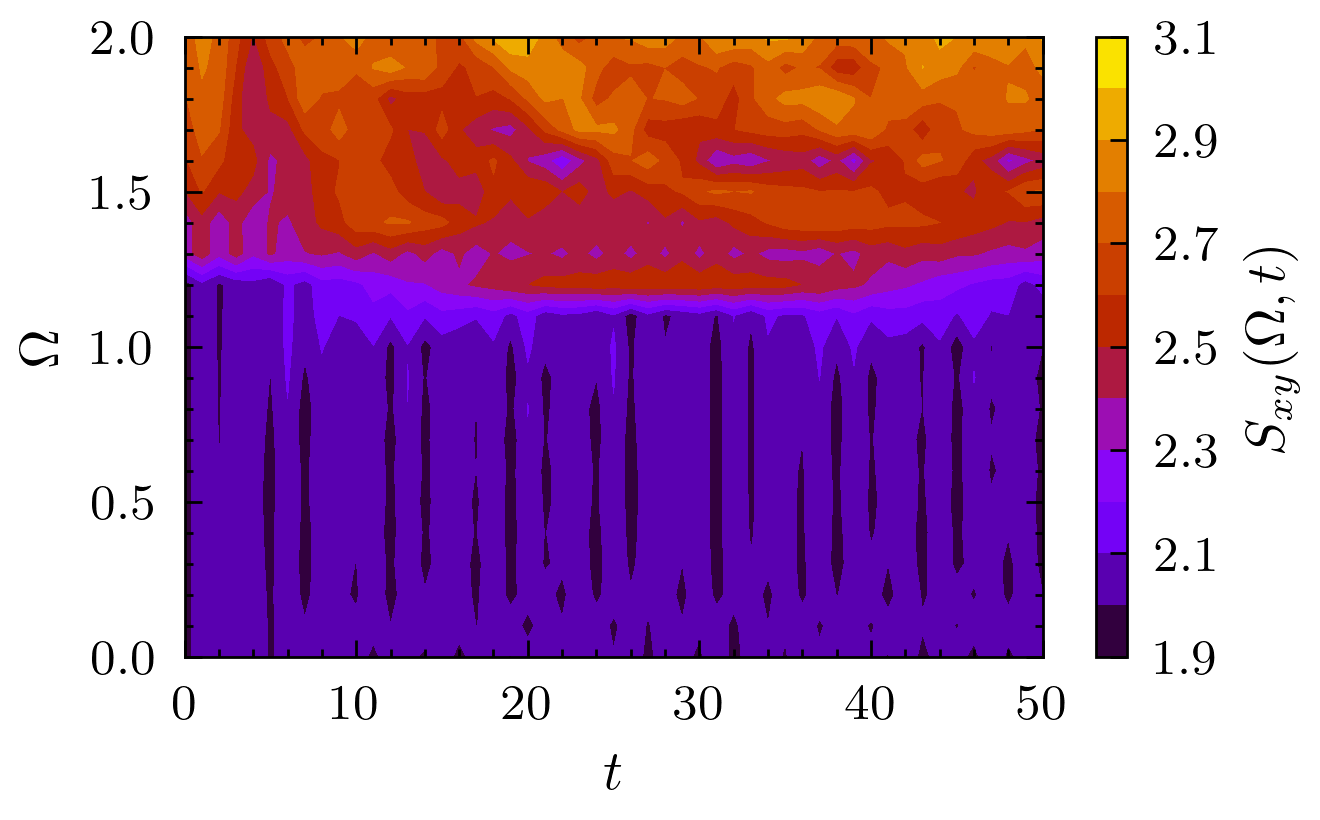}
    
    \caption{Phase diagrams of the marginal entropies $S_x$ (a), $S_y$  (b), and joint entropy $S_{xy}$ (c)  as a function of rotation frequency $\Omega$ and time $t$ following the trap quench. All three measures exhibit low, regular values for nonvortex states at small $\Omega$ 
and large, strongly fluctuating values in the high-$\Omega$ chaotic regime, confirming the qualitative features observed in the mutual information phase diagram.}
    \label{fig:three_in_column}
\end{figure}

To complement the mutual information phase diagram presented in the main
text, Fig.~\ref{fig:three_in_column} shows the dynamical maps of the
marginal entropies $S_x(t)$ and $S_y(t)$ and the joint entropy
$S_{xy}(t)$ as functions of rotation frequency and time following the
trap quench. All three quantities exhibit low-amplitude, regular
variations at small $\Omega$ and increasingly pronounced fluctuations
at larger $\Omega$, reflecting the progressive redistribution and
delocalization of the density.

Unlike the mutual information $I(t,\Omega)$, however, the individual
entropy measures do not clearly resolve the intermediate dynamical
regime. This difference follows from their physical content: $S_x$,
$S_y$, and $S_{xy}$ primarily characterize the spatial extent of the
density distribution, whereas the mutual information
$I=S_x+S_y-S_{xy}$ isolates the correlations between the two spatial
degrees of freedom. Consequently, $I(t,\Omega)$ provides greater
sensitivity to the intermediate regime, where correlations begin to
develop without a correspondingly strong change in the overall spatial
delocalization. At higher rotation frequencies, the enhanced magnitude
and temporal fluctuations of $I$ reflect the increasingly correlated
and irregular vortex dynamics.

\section{Numerical Convergence and Conservation Checks}

\subsection{Orbital-number convergence}

\begin{figure}[htbp]
    \centering
    \includegraphics[width=\columnwidth]{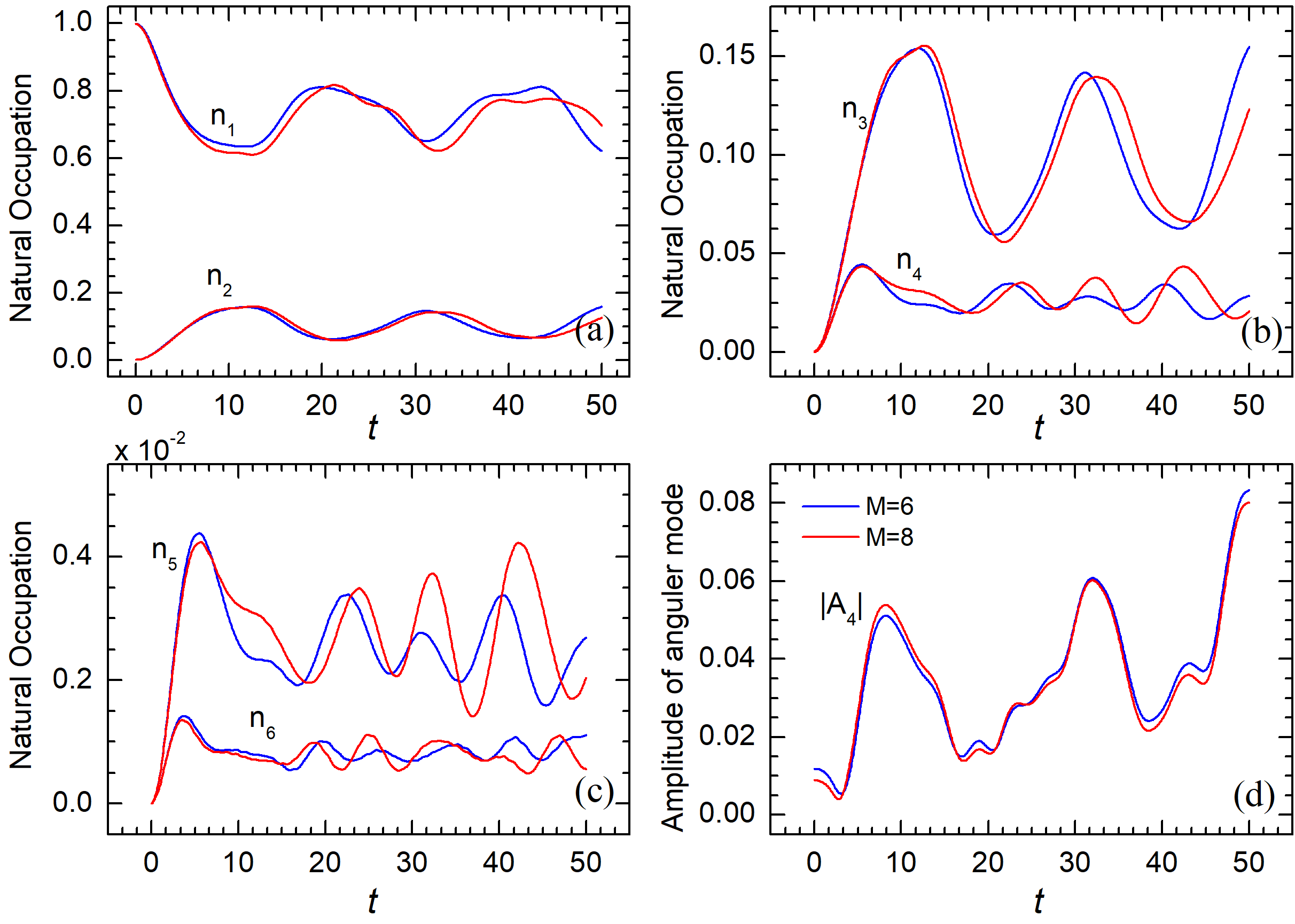}
    \caption{Orbital-number convergence for the strongly interacting quench at $\Omega=2.0$. The time evolution of the natural occupations obtained with $M=6$ and $M=8$ is shown for (a) $n_1$ and $n_2$, (b) $n_3$ and $n_4$, and (c) $n_5$ and $n_6$. The corresponding dynamics of the observable $A_4$ are shown in (d). The $M=6$ and $M=8$ results exhibit the same qualitative behavior, with only minor quantitative deviations, demonstrating convergence with respect to the number of orbitals.}
    \label{fig:orbit-convergence}
\end{figure}

To assess convergence with respect to the number of time-dependent
orbitals, we consider the strongly interacting quench at $\Omega=2.0$,
where dynamical many-body fragmentation is appreciable. The results
obtained with $M=6$, used for the main-text calculations, are compared
with those obtained using $M=8$. 
Fig.~\ref{fig:orbit-convergence}(a)--(c) shows the time evolution of the natural-orbital occupations, grouped as $(n_1,n_2)$, $(n_3,n_4)$, and $(n_5,n_6)$, respectively. The two most significantly occupied orbitals, $n_1$ and $n_2$, exhibit negligible differences between the $M=6$ and $M=8$ calculations and follow essentially the same temporal evolution. Small differences appear in the temporal evolution of the higher orbitals, while their overall contribution remains subdominant. These differences become more visible at later times, particularly in the occupations of the less populated orbitals, indicating some sensitivity of the detailed occupation dynamics to the orbital number. Nevertheless, the dominant natural-orbital populations remain in close agreement, supporting the overall convergence of the results.

As an additional convergence check, we compare the fourfold angular
observable $A_4(t)$ in Fig.~\ref{fig:orbit-convergence}(d). The $M=6$ and
$M=8$ calculations exhibit very similar temporal behavior, with only
minor differences in amplitude. The negligible variation of the
dominant occupations, together with the close agreement of the principal
angular observable $A_4(t)$, demonstrates that increasing the orbital
number from $M=6$ to $M=8$ does not alter the main dynamical features or
the physical conclusions of the calculation. We therefore consider
$M=6$ sufficient for the present study.

\subsection{Spatial-grid convergence}

\begin{figure}[htbp]
    \centering
    \includegraphics[width=\columnwidth]{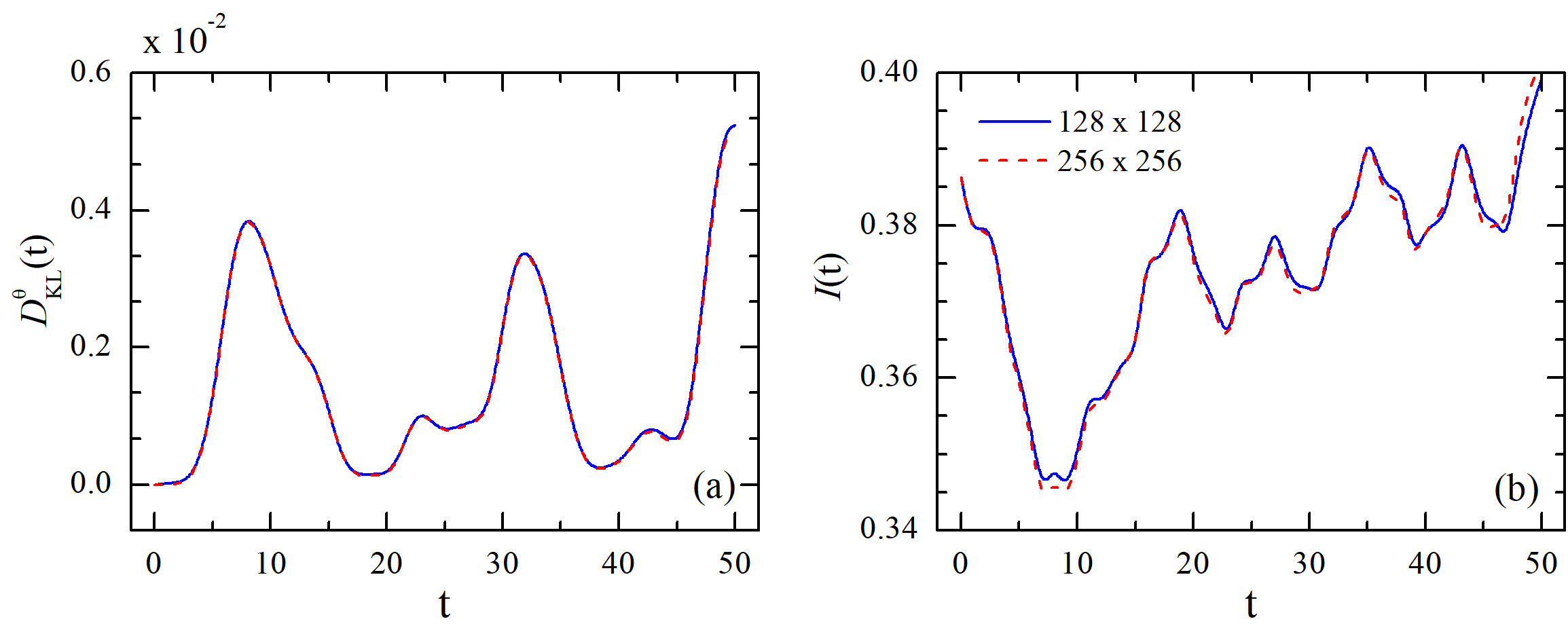}
  \caption{Spatial-grid convergence for the strongly interacting quench at
$\Omega=2.0$. The time evolution of (a) the angular
Kullback--Leibler divergence $D_{\mathrm{KL}}^{\theta}(t)$ and (b) the
mutual information $I(t)$ is shown for $128\times128$ and
$256\times256$ spatial grids. The two results are indistinguishable,
justifying the use of the $128\times128$ grid in the main text.}
    \label{fig:grid-convergence}
\end{figure}

To assess convergence with respect to the spatial discretization, we
compare calculations performed on $128\times128$ and $256\times256$
grids for the strongly interacting quench at $\Omega=2.0$, keeping all
other parameters fixed. Fig.~\ref{fig:grid-convergence} compares the
time-dependent angular Kullback--Leibler divergence
$D_{\mathrm{KL}}^{\theta}(t)$ [panel (a)] and mutual information
$I(t)$ [panel (b)] at the two grid resolutions. The results obtained
with the two grid sizes are indistinguishable on the scale of the plots,
confirming that the spatial discretization does not affect the reported
dynamics. This validates the use of $128\times128$ grid points in the
main text as sufficient for the present calculations.

\subsection{Time-step convergence}

\begin{figure}[htbp]
    \centering
    \includegraphics[width=\columnwidth]{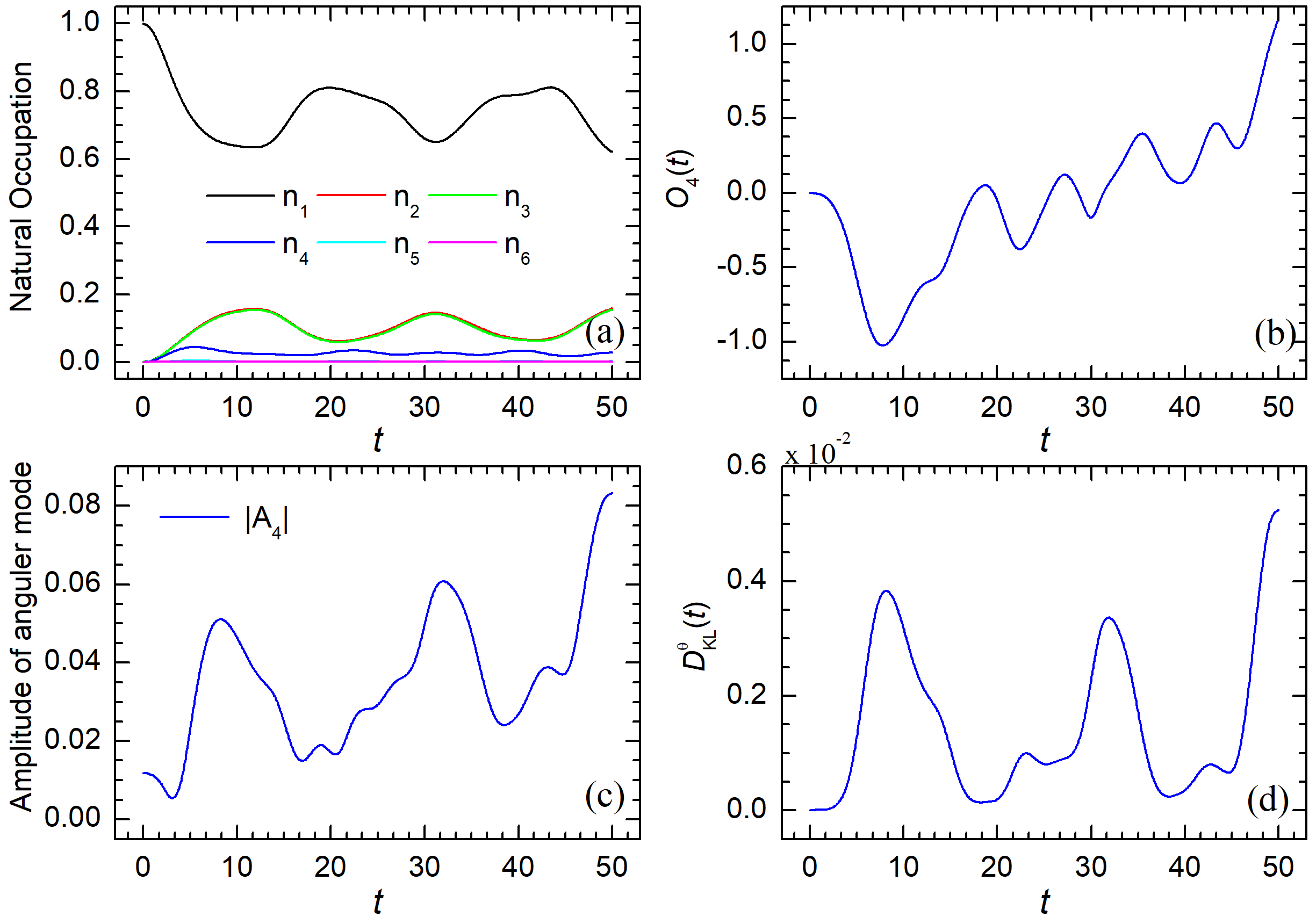}
  \caption{Time-step convergence of the main dynamical observables for the
strongly interacting quench. Results obtained using the production time
step and a reduced time step are compared for (a) the natural orbital
occupation, (b) the fourfold operator $O_4(t)$, (c) the angular Fourier
component $A_4(t)$, and (d) the angular Kullback--Leibler divergence
$D_{\mathrm{KL}}^{\theta}(t)$. The two calculations are indistinguishable,
confirming convergence of the reported dynamics with respect to the
time-step size.}
    \label{fig:time-convergence}
\end{figure}

To assess convergence with respect to the time-step size, we repeat the
calculation corresponding to Fig.~6 of the main text using a reduced
integration time step, $\Delta t_{\rm red} =
\Delta t/10= 0.001$, while keeping
all other simulation parameters unchanged. The resulting observables
are shown in Fig.~\ref{fig:time-convergence}. The results obtained with
the reduced time step are indistinguishable from those presented in
Fig.~6 of the main text, with no noticeable deviations in the temporal
evolution of the natural orbital occupation, $O_4(t)$, $A_4(t)$, or
$D_{\mathrm{KL}}^{\theta}(t)$. This confirms that the reported dynamics
are well converged with respect to the time-step size and are not
affected by temporal discretization.

\subsection{Conservation diagnostics}

\begin{figure}[htbp]
    \centering
    \includegraphics[width=\columnwidth]{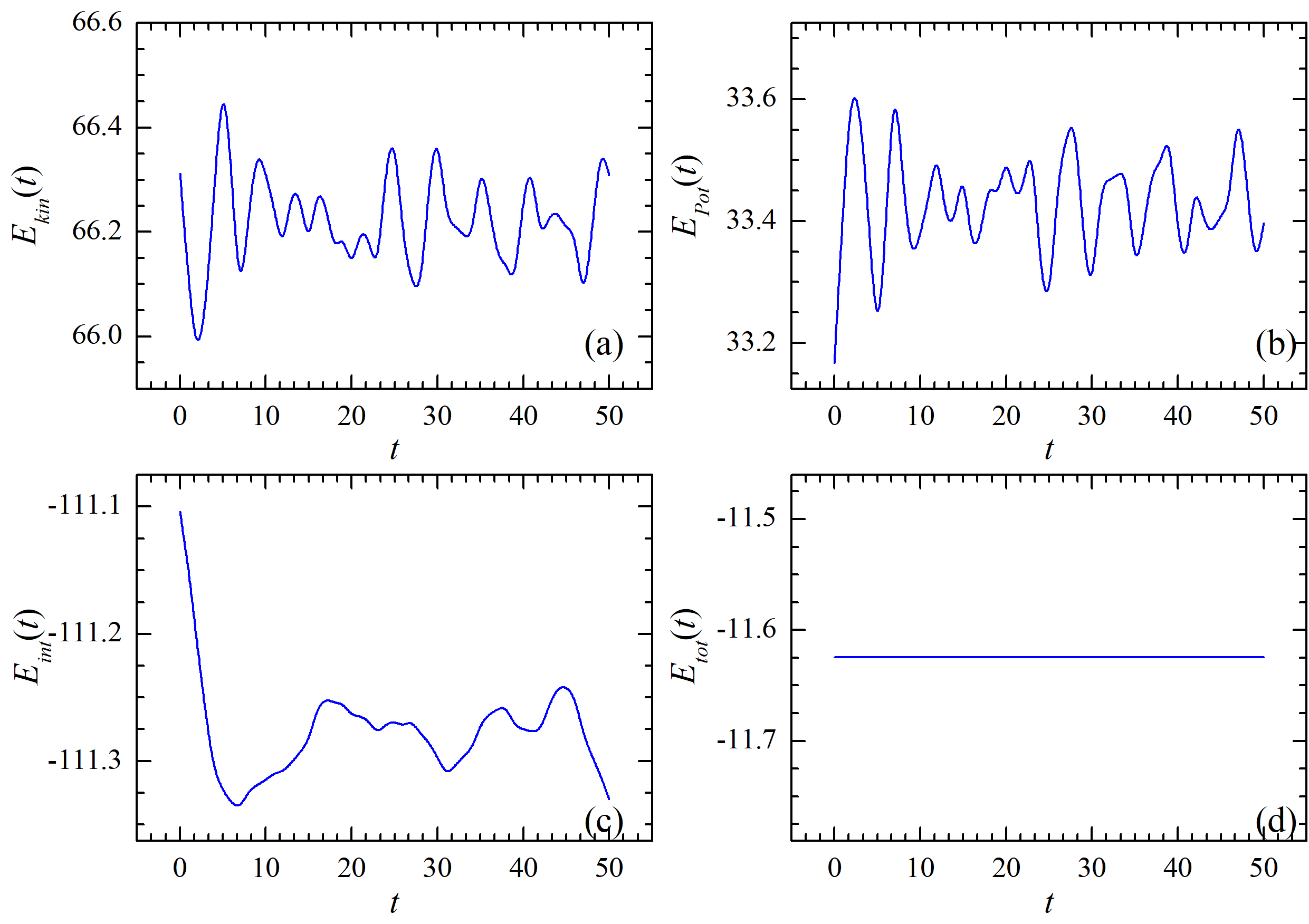}
  \caption{Energy conservation during the real-time evolution for the
strongly interacting quench at $\Omega=2.0$. The time evolution of
(a) the kinetic energy $E_{\mathrm{kin}}$, (b) the potential energy
$E_{\mathrm{pot}}$, (c) the interaction energy $E_{\mathrm{int}}$, and
(d) the total energy $E_{\mathrm{tot}}$ is shown. While the kinetic and
potential energies exhibit pronounced oscillations and the interaction
energy remains nearly unchanged, the total energy remains constant
throughout the propagation, confirming energy conservation and the
numerical stability of the time evolution.}
    \label{fig:diagnostic}
\end{figure}

The numerical stability of the real-time propagation is further assessed
for the strongly interacting quench at $\Omega=2.0$. Fig.~\ref{fig:diagnostic}
shows the time evolution of (a) the kinetic energy $E_{\mathrm{kin}}$,
(b) the potential energy $E_{\mathrm{pot}}$, (c) the interaction energy
$E_{\mathrm{int}}$, and (d) the total energy $E_{\mathrm{tot}}$. The
kinetic and potential energies exhibit pronounced oscillations,
reflecting the redistribution of energy among the different
contributions during the dynamics, while the interaction energy remains
comparatively stable. Despite the variations in the individual energy
components, the total energy remains constant throughout the
propagation. This conservation confirms the numerical stability of the
time evolution and supports the reliability of the reported
nonequilibrium dynamics.

\section{Particle-number scaling of the giant-vortex instability}

To assess the particle-number dependence of the observed four-fold giant-vortex
instability, we examine the evolution of the $A_4$ mode under two different
interaction-scaling protocols.

\begin{figure}[htbp]
    \centering
    \includegraphics[width=\columnwidth]{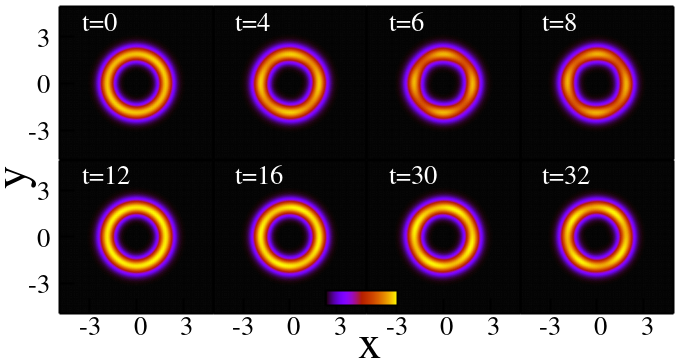}
    \caption{Time evolution of the density for the $\Omega=2.0$ condensate following a stronger interaction quench ($g^{\prime} \approx 5 g$) with particle number $N=16$. The central giant vortex (dark circular region) remains largely circular, while the bright outer ring develops a fourfold modulation as observed in Fig.~\ref{fig:mono-density-20}, splitting into four partially connected high-density segments.}
    \label{fig:mono-density-20-N=16}
\end{figure}

First, we increase the particle number while keeping the microscopic
interaction strength $g$ fixed. Within the particle numbers accessible to
our simulations, the four-fold instability persists upon increasing $N$
from the value $N=8$ used in the main calculations. In particular, the
calculation at $N=16$ still exhibits a pronounced four-fold deformation of
the giant-vortex state as shown in Fig.~\ref{fig:mono-density-20-N=16}. We reduced number of orbitals to $M=4$. Because of the rapidly increasing computational
cost with particle number, we do not extend the fixed-$g$ calculations to
substantially larger systems.

We next consider the conventional mean-field scaling, in which the effective interaction parameter \(\Lambda=g(N-1)\) is kept constant at \(\Lambda=0.1\), corresponding to \(g=0.1/(N-1)\). Accordingly, the interaction strength decreases with increasing particle number, taking the values \(g=0.014286\), \(0.006667\), \(0.002041\), and \(0.001010\) for \(N=8\), \(16\), \(50\), and \(100\), respectively. We use \(M=6\) orbitals for \(N=8\) and \(M=4\) orbitals for the larger-particle-number calculations (\(N=16,50,100\)).

\begin{figure}[htbp]
    \centering
    \includegraphics[width=\columnwidth]{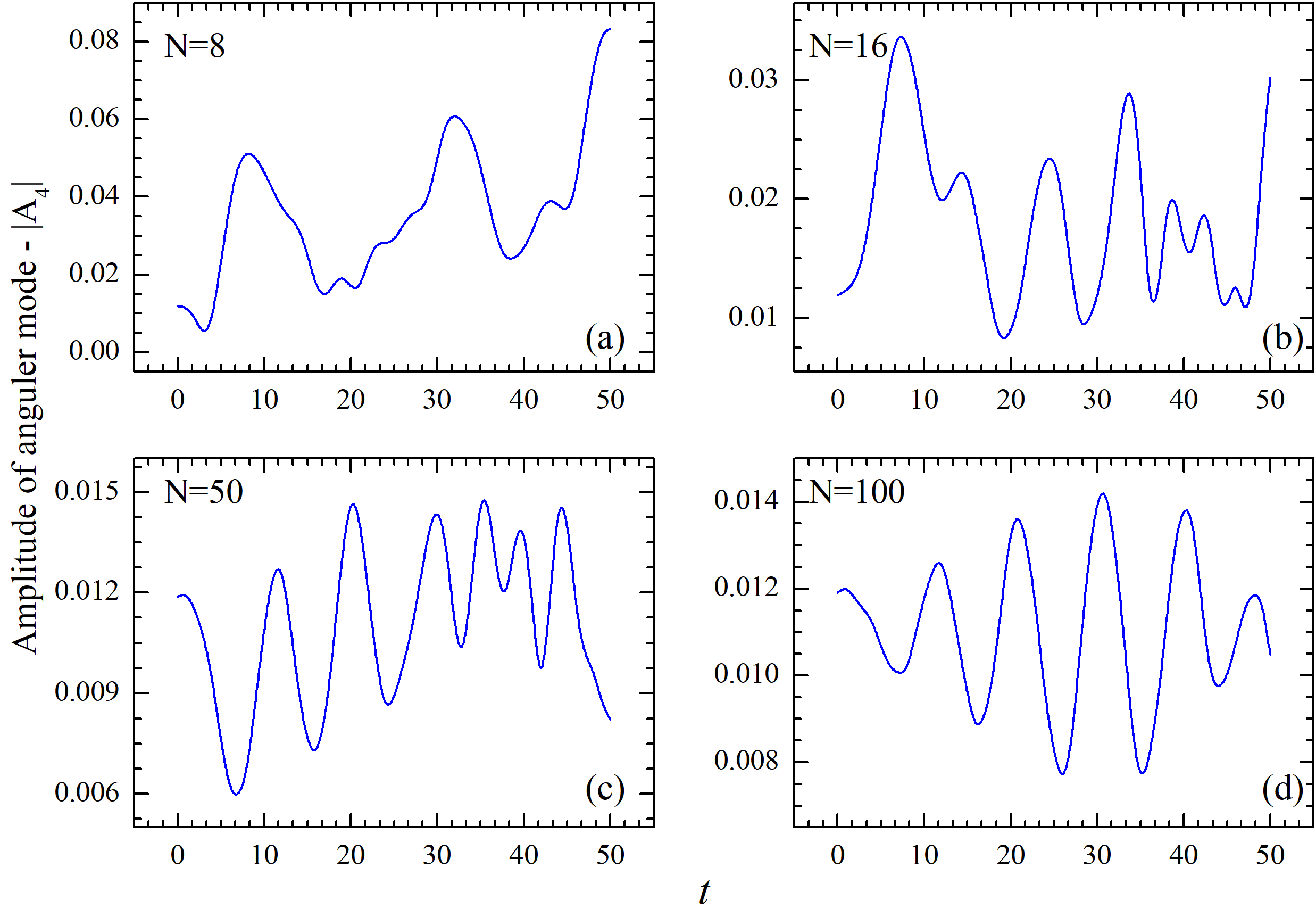}
\caption{Particle-number dependence of the four-fold anisotropy $A_4$
under the mean-field scaling $Ng=\mathrm{const.}$. The time evolution of
$A_4$ is shown for $N=8$, $16$, $50$, and $100$. While the four-fold
response remains finite for all particle numbers considered, its
amplitude decreases with increasing $N$, and the dynamics become
increasingly oscillatory, indicating a progressive suppression of the
four-fold instability with increasing particle number under mean-field
scaling.}
    \label{fig:mean-field-scaling}
\end{figure}

Fig.~\ref{fig:mean-field-scaling} shows the time evolution of the four-fold
anisotropy $A_4$ for $N=8$, $16$, $50$, and $100$. The $A_4$ response remains
finite for all particle numbers considered, indicating that the four-fold
deformation persists under this scaling. However, its amplitude decreases
systematically with increasing $N$, while the dynamics become increasingly
regular and oscillatory.

Thus, the particle-number dependence of the giant-vortex dynamics depends
on the interaction-scaling protocol. At fixed $g$, the four-fold instability
remains clearly observable over the accessible range of particle numbers.
Under the mean-field scaling $Ng=\mathrm{const.}$, its amplitude decreases
with increasing $N$, indicating a weakening of the four-fold response as
the system approaches the mean-field limit. Nevertheless, a finite
four-fold response persists over the range of particle numbers studied.

\begin{figure}[htbp]
    \centering
    \includegraphics[width=\columnwidth]{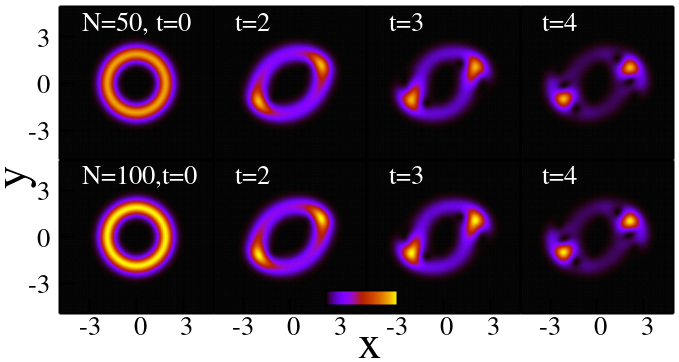}
\caption{Density snapshots illustrating the particle-number dependence
of the giant-vortex instability following the trap quench. The upper and
lower panels correspond to $N=50$ and $N=100$, respectively. In both
cases, the characteristic giant-vortex instability and subsequent
deformation of the density distribution persist, consistent with the
behavior observed for $N=8$ in the main text.}
    \label{fig:particle-quad}
\end{figure}

To further examine the particle-number dependence of the giant-vortex
instability, we performed trap-quench simulations for larger particle
numbers, $N=50$ and $N=100$, $M=4$. The corresponding density snapshots are
shown in Fig.~\ref{fig:particle-quad}, with $N=50$ and $N=100$ displayed
in the upper and lower panels, respectively. Despite the substantial
increase in particle number, both cases exhibit the characteristic
giant-vortex instability and subsequent deformation of the density
distribution observed for $N=8$ in the main text. These results provide
additional evidence that the underlying instability persists at
substantially larger particle numbers. The reduced numerical cost of the
trap-quench protocol allows us to access these larger-$N$ systems
directly.

\section{Information-theoretic measures as quantitative diagnostics of vortex dynamics}

\begin{figure}[htbp]
    \centering
    \includegraphics[width=\columnwidth]{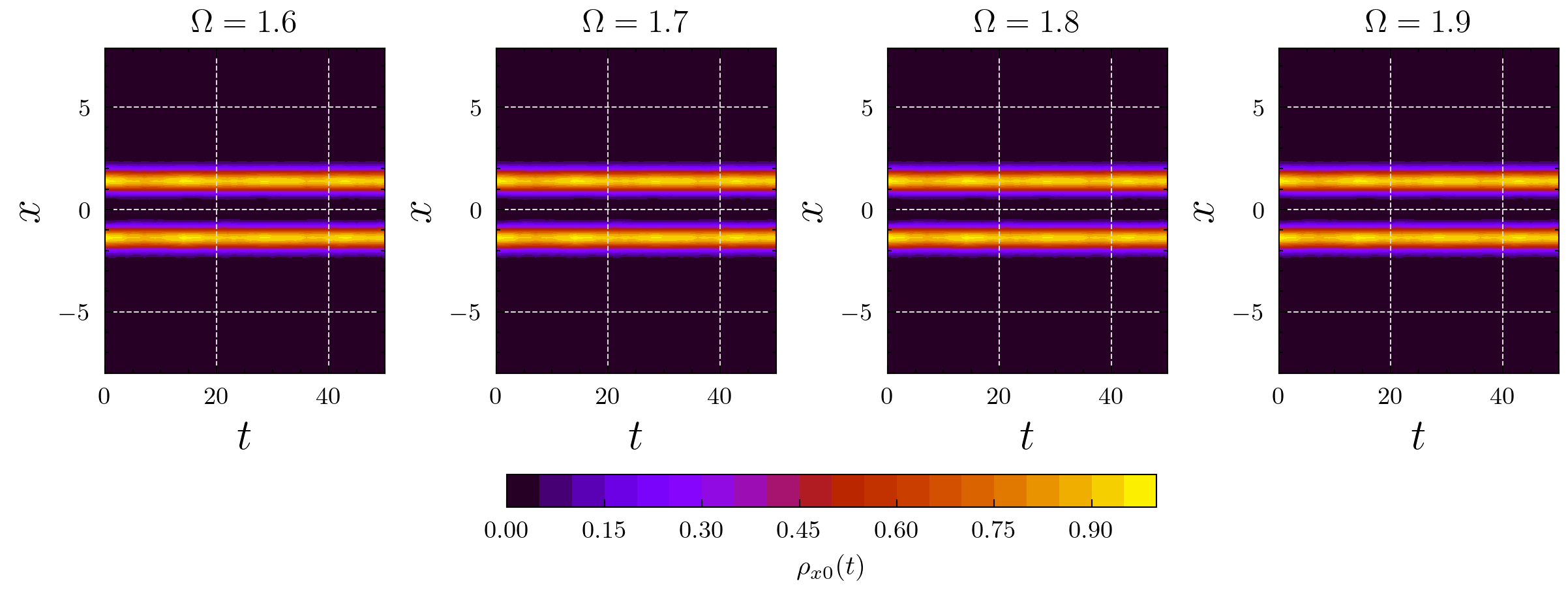}
\caption{Time evolution of the density along the central slice, $\rho(x,0,t)$, following strong interaction quench, for rotation frequencies $\Omega=1.6$, $1.7$, $1.8$, and $1.9$. The two high-density structures delimiting the central depleted region become progressively more separated with increasing $\Omega$, indicating an increase in the characteristic spatial extent of the giant-vortex core. At larger rotation frequencies, the density structures also exhibit stronger temporal deformations.}
    \label{fig:rhox0}
\end{figure}

\begin{figure}[htbp]
    \centering
    \includegraphics[width=\columnwidth]{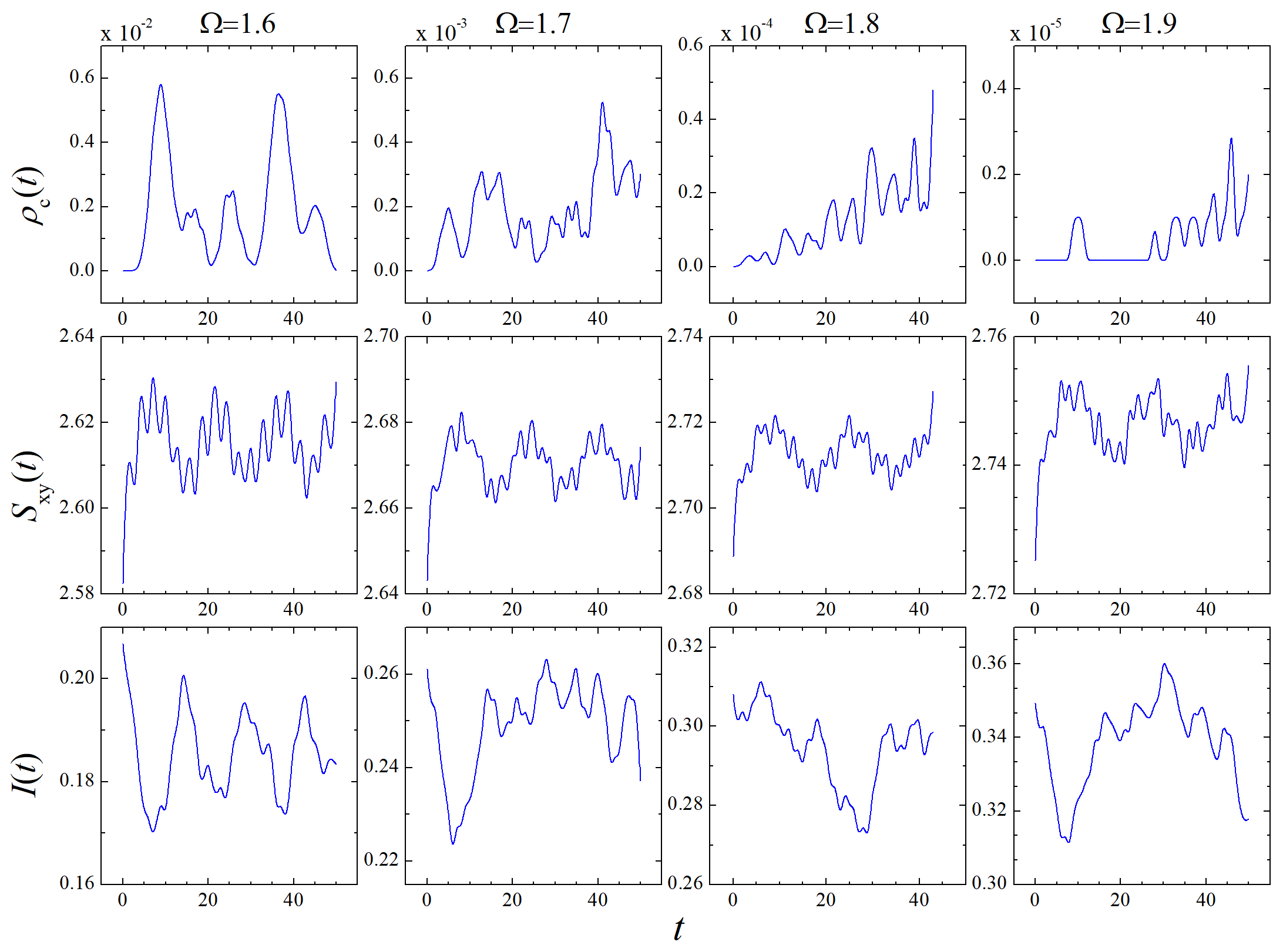}
\caption{Comparison of conventional and information-theoretic measures, following strong interaction quench, for rotation frequencies $\Omega=1.6$, $1.7$, $1.8$, and $1.9$. The top row shows the central density $\rho_c(t)=\rho(0,t)$, which quantifies the degree of central depletion. The middle and bottom rows show the spatial entropy $S_{xy}(t)$ and mutual information $I(t)$, respectively. While $\rho_c(t)$ is progressively suppressed with increasing rotation frequency, indicating stronger central depletion, $S_{xy}(t)$ and $I(t)$ exhibit temporal fluctuations with an overall increase in magnitude, providing complementary global measures of the spatial information and correlation structure generated during the vortex dynamics.}
    \label{fig:IT}
\end{figure}

To further clarify the physical information provided by the IT measures, we compare them with direct
density-based diagnostics of the vortex dynamics. Fig.~\ref{fig:rhox0}
shows the time evolution of the density along the central slice,
$\rho(x,0,t)$, for $\Omega=1.6$, $1.7$, $1.8$, and $1.9$. At
$\Omega=1.6$, two well-defined high-density structures delimit the
central depleted region and remain approximately stationary throughout
the evolution. Their separation increases with increasing rotation
frequency, indicating an increasing spatial extent of the central
giant-vortex region. At larger rotation frequencies, particularly
$\Omega=1.9$, the density structures exhibit stronger
temporal deformations. While these density maps provide a direct
visualization of the vortex evolution, extracting a single quantitative
measure of the increasingly complex temporal deformation from the
density profiles is not straightforward.

To quantify the evolving dynamics, we compare the central density $\rho_c(t)=\rho(0,t)$ with the IT measures $S_{xy}(t)$ and $I(t)$ in
Fig.~\ref{fig:IT}. The central density exhibits temporal oscillations,
while its overall magnitude decreases markedly with increasing rotation
frequency. At lower rotation frequencies, $\rho_c(t)$ remains of order
$10^{-3}$, whereas at $\Omega=1.9$ it is suppressed to values of order
$10^{-6}$. Thus, $\rho_c(t)$ provides a direct quantitative measure of
the depth of the central density depletion associated with the
giant-vortex state. However, as the central density approaches near-complete depletion, its absolute temporal variation becomes increasingly small, 
making $\rho_c(t)$ less sensitive to further redistribution of density away from the core. In contrast, $S_{xy}$
and $I$ remain appreciable and continue to exhibit systematic temporal variations, reflecting global changes in the spatial distribution and correlations even when the core density is nearly depleted.

This comparison highlights the complementary character of the IT
measures. The central density $\rho_c(t)$ probes a specific local
property of the density, namely the depth of the central depletion,
whereas $S_{xy}(t)$ quantifies the global spatial spreading of the
density and $I(t)$ measures correlations between the $x$ and $y$
degrees of freedom. Consequently, the IT measures provide quantitative
information beyond the central-density diagnostic alone. In particular,
their continued variation with rotation frequency provides a measure
of changes in the global spatial distribution and inter-directional
correlations even when the central density becomes strongly suppressed and insensitive.

The IT measures should therefore be regarded as complementary rather
than superior to conventional density-based observables. The density
profiles and $\rho_c(t)$ directly characterize the local structure and
depth of the vortex core, while the entropy and mutual information
quantify global spatial redistribution and correlations. Their combined
use provides a more complete quantitative characterization of the
vortex dynamics, connecting the local density evolution with the global
information and correlation structure of the evolving state.

\bibliography{ref}

\end{document}